%% file: main.tex
\documentclass[manuscript,screen]{acmart}

\AtBeginDocument{%
  \providecommand\BibTeX{{%
    \normalfont B\kern-0.5em{\scshape i\kern-0.25em b}\kern-0.8em\TeX}}}

\setcopyright{acmcopyright}
\copyrightyear{2018}
\acmYear{2018}
\acmDOI{XXXXXXX.XXXXXXX}

\acmConference[Conference acronym 'XX]{Make sure to enter the correct
  conference title from your rights confirmation email}{June 03--05,
  2018}{Woodstock, NY}
\acmPrice{15.00}
\acmISBN{978-1-4503-XXXX-X/18/06}

\usepackage{wrapfig}
\usepackage{amsmath,amsfonts}
\usepackage[inkscapelatex=false]{svg}
\usepackage{enumitem}
\usepackage{caption, booktabs}
\usepackage{subcaption}
\usepackage{booktabs}
\usepackage{lscape}
\usepackage{placeins}
\usepackage{textcomp}
\usepackage{multirow}
\usepackage{array}
\usepackage{listings}
\usepackage{color}
\usepackage{changepage}
\usepackage{tikz}
\usepackage{algorithm,algpseudocode}
\usepackage{url}

\begin{document}
% \sloppy
%\title{Judging the Team by its Cover: A Study on the Evolution of Concepts in Distributed Development Teams}

\title[Balanced Knowledge Distribution among Software Development Teams]{Balanced Knowledge Distribution among Software Development Teams - Observations from Open-Source and Closed-Source Software Development}

\author{Saad Shafiq}
\email{saad.shafiq@jku.at}
\orcid{0000-0002-5901-1420}
\affiliation{%
  \institution{Johannes Kepler University}
  \city{Linz}
  \country{Austria}
}

\author{Christoph Mayr-Dorn}
\affiliation{%
  \institution{Johannes Kepler University}
  \city{Linz}
  \country{Austria}}
\email{christoph.mayr-dorn@jku.at}

\author{Atif Mashkoor}
\affiliation{%
  \institution{Johannes Kepler University}
  \city{Linz}
  \country{Austria}
}
\email{atif.mashkoor@jku.at}

\author{Alexander Egyed}
\affiliation{%
 \institution{Johannes Kepler University}
 \city{Linz}
 \country{Austria}}
\email{alexander.egyed@jku.at}

\renewcommand{\shortauthors}{Shafiq, et al.}

\begin{abstract}

In software development teams, developer turnover is among the primary reasons for project failures as it leads to a great void of knowledge and strain for the newcomers. Unfortunately, no established methods exist to measure how knowledge is distributed among development teams. Knowing how this knowledge evolves and is owned by key developers in a project helps managers reduce risks caused by turnover. To this end, this paper introduces a novel, realistic representation of domain knowledge distribution: the \textit{ConceptRealm}. To construct the \textit{ConceptRealm}, we employ a latent Dirichlet allocation model to represent textual features obtained from 300k issues and 1.3M comments from 518 open-source projects. We analyze whether the newly emerged issues and developers share similar concepts or how aligned the developers' concepts are with the team over time. We also investigate the impact of leaving members on the frequency of concepts. Finally, we evaluate the soundness of our approach to closed-source software, thus allowing the validation of the results from a practical standpoint. We find out that the \textit{ConceptRealm} can represent the high-level domain knowledge within a team and can be utilized to predict the alignment of developers with issues. We also observe that projects exhibit many keepers independent of project maturity and that abruptly leaving keepers harm the team's concept familiarity.

% The turnover of developers has detrimental long-term implications in software development that vary from loss of critical knowledge leading to retraining of new comers and poor product quality. It has steered many organizations into transitioning towards adapting new strategies in order to cope with such loss. Researchers have explored the interaction of core developers with the software artifacts in the past, however, in essence, the evolution of concepts emerging within advanced software development teams still remains unaddressed. 
% Understanding this evolution of a software project involves knowing the association of team concepts with the project stakeholders and how they change overtime. In this study, we aim to extend our previous work, we extract team concepts that evolve over the course of iterations and identify the knowledge distributions among projects with the help of these concepts. It would further help us understand whether core developers are assigned to issues more in equally distributed projects or otherwise. Ultimately allowing us to provide recommendations which ensure a balanced distribution of knowledge within teams.
% We also evaluated our approach in an industrial setting, allowing us to better validate the results from the practical standpoint and also to be able to compare across OSS systems.

\end{abstract}

% \begin{IEEEkeywords}
% Software evolution, Developer knowledge, Software engineering 
% \end{IEEEkeywords}

%   <concept>
%       <concept_id>10011007.10011074.10011134</concept_id>
%       <concept_desc>Software and its engineering~Collaboration in software development</concept_desc>
%       <concept_significance>500</concept_significance>
%       </concept>
%       <concept_id>10010147.10010178.10010179</concept_id>
%       <concept_desc>Computing methodologies~Natural language processing</concept_desc>
%       <concept_significance>500</concept_significance>
%       </concept>
%     <concept>
%       <concept_id>10011007.10011074.10011081.10011091</concept_id>
%       <concept_desc>Software and its engineering~Risk management</concept_desc>
%       <concept_significance>300</concept_significance>
%       </concept>
%   <concept>
%  </ccs2012>

% \ccsdesc[500]{Software and its engineering~Collaboration in software development}
% \ccsdesc[500]{Computing methodologies~Natural language processing}
% \ccsdesc[300]{Software and its engineering~Risk management}

\keywords{software evolution, developer knowledge, software engineering}

\maketitle

\section{Introduction}
\label{sec:intro}
\input{intro.tex}

\section{Background and Related Work}
\label{sec:background}
\input{background.tex}

\section{Motivation}
\label{sec:motivation}
\input{motivation.tex}

\section{Defining the \textit{ConceptRealm}}
\label{sec:definition}

\input{definition.tex}

\section{Study Design}
\label{sec:studydesign}
\input{studydesign.tex}

\section{Results}
\label{sec:results}
\input{results.tex}

\section{Discussion and implications}
\label{sec:discussion}
\input{discussion.tex}

\section{Threats to validity}
\label{sec:ttv}
\input{ThreatsToValidity.tex}

\section{Conclusion}
\label{sec:conclusion}
\input{conclusion.tex}

% \section{Approach}
% \label{sec:approach}
% \input{approach.tex}

% \section{Algorithm}
% \label{sec:algo}
% \input{algorithm.tex}

% \section{Evaluation}
% \label{sec:evaluation}
% \input{evaluation.tex}

%\section*{Acknowledgment}

%The research reported in this paper has been partly funded by the Linz Institute of Technology, and the Austrian Federal Ministry of Transport, Innovation and Technology, the Austrian Federal Ministry for Digital and Economic Affairs and the Province of Upper Austria and Styria in the frame of the COMET Program managed by FFG.

\begin{acks}

The research reported in this paper has been partly funded by the Linz Institute of Technology, and the Austrian Federal Ministry of Transport, Innovation and Technology, the Austrian Federal Ministry for Digital and Economic Affairs and the Province of Upper Austria and Styria in the frame of the COMET Program managed by FFG. We would also like to acknowledge the support from Philipp Lengauer at Dynatrace for providing us with the data and helping us with the evaluation.

\end{acks}

\bibliographystyle{ACM-Reference-Format}
\bibliography{BalancedDist.bib}

\end{document}

%% file: intro.tex
Among the prime reasons for the failure of software projects are high budget costs and time overruns~\cite{Sylvester2012}. One of the causes of these failures is the underestimation of employee turnover rate~\cite{Boehm2007,Cosentino2015}: (core) developers leaving and new developers joining for various reasons~\cite{Xuan2012,Robillard2021} resulting in project delays. Many companies now involve risk managers to overcome the risk of such failures~\cite{McManus2012, Lin2017}. The critical aspect of turnover is the loss of knowledge~\cite{DeMelo2013}. This is especially relevant in open-source software (OSS) development, which is much more prone to developer turnover than closed-source projects~\cite{Foucault2015} and hence is more prone to an imbalanced distribution of knowledge across team members.

One essential element in making this risk more manageable is identifying keepers, i.e., core developers of a team possessing essential knowledge related to the project. To characterize the knowledge possessed by the team regarding various artifacts, the term ``concept'' has been employed in the literature~\cite{Abebe2010,Abebe2015}. A concept represents the knowledge regarding the domain and implementation of the software project. Similar to topic modeling~\cite{Liao2017}, a concept is a collection of terms that are highly similar and coherent to each other (see Section~\ref{CE} for a more detailed definition and discussion). 

It is generally not well understood to what extent concepts are distributed in software development teams. Several researchers have extracted concepts from artifacts such as source code but hardly attempted to link concepts to developers. One such prior work that focuses on the developer-centric concept is by Dey et al.~\cite{Dey2020} who analyzed developer familiarity specific to using third-party libraries in source code. Their approach thus describes a skill set in lower-level libraries. In contrast, our approach focuses on establishing the concepts that describe the actual software system under development.
To provide developers and team leads with tools that help, for example, to manage and assess the risk of developer turnover or identify developers with particular expertise, we need not only to determine existing concepts and their distribution in the team but also how this distribution changes over time. This allows assessing whether certain concepts are still relevant, whether concept distribution measures to reduce turnover risk indeed have the expected effect, and whether new concepts need to be considered.

% New paper intro

Events such as developer turnover can cause projects high costs. These costs include hiring and training of newcomers~\cite{Mockus2010}, resulting in poor product quality and delays in software projects. Furthermore, with core members leaving the projects, there is also a substantial loss of important tacit knowledge~\cite{Mockus2009,Mockus2010,Yuan2011}.
The organizations have employed various strategies to reduce such losses, e.g., maintaining updated documentation and enabling active learning sessions within teams~\cite{Nidhra2013}. However, capturing the tacit knowledge owned by the core developers is still an ongoing research topic.

To this end, this paper aims to construct a novel representation -- called the~\textit{ConceptRealm} -- of domain knowledge distribution in a team. We treat the~\textit{ConceptRealm} as a representation of the problem space (as captured in issues) and the corresponding developers' familiarity with it. We build the~\textit{ConceptRealm} from the textual features obtained from issues, and their comments as these reflect the focus of the team better than slowly changing artifacts such as documentation or requirements. In addition, we examine the distribution of concepts among projects and the reasons for the differences that exist among them. Lastly, we aim to use~\textit{ConceptRealm} to derive implications for recommendation algorithms to balance knowledge within the team. We evaluate this representation on OSS projects and a field study on the closed-source project.

In this work, we introduce a novel representation to capture this tacit knowledge with the help of concepts. Applying our approach to more than 500 open-source projects comprising more than 300k issues and over a million comments, we determined that we can extract meaningful concepts that allow identifying concept keepers and that when such keepers leave the project, a drop in that concept's familiarity within the team is likely to follow. We also find that most investigated open source projects exhibit keepers for their most essential concepts regardless of project age. 

To determine the relevance of results from a practical standpoint, we have evaluated it using an industrial case study from Dynatrace\footnote{\url{https://www.dynatrace.com/}}. Dynatrace is a product-based company providing organizations with a diverse software intelligence platform promoting observability and infrastructure to facilitate cloud automation in their systems. This project consists of 49457 issues and 25807 comments in total. We extracted the concepts from this project in a similar way as from OSS projects. However, we also had access to the teams assigned to the issues in this project which further strengthened our findings.
The industrial validation of our results attests that the~\textit{ConceptRealm} is a practical approach for supporting stakeholders that need to manage the risk of developer turnover by aiming for balanced concept distribution. 

% STUDY OUTCOME:
The contributions of this article are six-fold: 

\begin{itemize}
    
    \item a novel representation called \textit{ConceptRealm} to capture the high-level domain knowledge and its distribution across team members.
    
    \item analysis of the fluctuations of concepts throughout an (ongoing) project's lifetime. 
    
    \item an approach to compare the keeping extent of developers across projects and monitor the impact when they leave\footnote{For simplicity, we will refer to developers that become less engaged with the project as leaving members through the rest of the paper.} the project.
    
    \item an industrial case study to demonstrate the practicality and usefulness of the proposed approach.
    
    \item a dataset consisting of concepts extracted from issues and comments and how strongly these issues and comments belong to a particular concept.
    
    \item an in-depth analysis of distribution differences among open- and closed-source projects.
    
 \end{itemize}

The remainder of the article is organized as follows: Section~\ref{sec:background} discusses the related work. The motivation of this study is explained in Section~\ref{sec:motivation}. The \textit{ConceptRealm} and corresponding metrics are introduced in Section~\ref{sec:definition}. Section~\ref{sec:studydesign} describes the evaluation methodology employed in this study. We provide results in Section~\ref{sec:results}. We discuss these results and their implications in Section~\ref{sec:discussion}. Section \ref{sec:ttv} discusses the threats to validity of this research. Lastly, we conclude this article with an outlook on future work in Section~\ref{sec:conclusion}.

%% file: background.tex
In this section, we describe the related work on the representation of developers' knowledge complemented with an elaboration on the word representations and topic modeling techniques that are considered influential in capturing the context of OSS development in literature.

\subsection{Socio-technical factors}

The geographically distributed nature of OSS projects has changed the standard practices of traditional software engineering. In OSS projects, a large number of contributors voluntarily take on the tasks of their own accord. In addition, due to the scattered locations of these contributors, they mostly rely on means of digital communication instead of meeting face to face in a collocated space. As Conway~\cite{Conway1968} said, designs are copies of the communication structure of the organization. The interest in the existence of socio-technical factors in OSS projects developed in the early days when researchers started analyzing the social aspects of OSS projects  such as electronic means of communication along with code writing~\cite{DILAN2002,Sack2006}. Since then, many studies have been conducted on investigating the effects of socio-technical factors on pull request quality in an OSS development environment~\cite{Dey2020b,Dey2020a,Tikhonov2014}. 

Researchers have also studied collaboration from various perspectives such as exploring team distribution, and communication patterns~\cite{Abdullah2011a,Ehrlich2014a,Panichella2014,Shafiq2017,Ortu2018,Shafiq2019,Kakimoto2006}. 
A study conducted by Von Krogh et al.~\cite{VonKrogh2003} focused on identifying the communication patterns that appear in the new contributors joining the project. The authors have called this a ``joining script'', which implies how much expertise newcomers require before being given access to make contributions to the projects. However, the study has made further research implications to explore the evolution of developers' expertise alignment with the expertise of the newcomers joining the project.

Ducheneaut et al.~\cite{Ducheneaut2005} performed an ethnographic study on the interaction between the OSS mailing lists and the codebases of OSS repositories in order to understand the solitary learning process of an individual and also the political process such as forming allies in the development process. The study concluded that a successful contribution goes beyond technical expertise. For individuals, defining their identity in OSS eventually leads them in becoming contributors to the project. Apart from this, OSS is also entangled with a political aspect, which refers to the opaque connections in the network necessary to sustain the project. New contributors must understand the black-box nature of connections in order to form allies that could support them in their contributions.

Panichella et al.~\cite{Panichella2014} investigated evolving collaboration via three communication channels including mailing lists, issue trackers, and IRC logs. The goal of this study was to determine whether the interaction of developers over social network impact code changes. The study was evaluated using seven open-source projects. Results showed that developers tend to use two out of three channels to communicate and an intense chat does not essentially correlate with high code impact.

Wu et al.~\cite{Wu2009} studied the effects of centrality measures on OSS projects. In particular, social network analysis measures, such as project density and leadership centrality, were used to evaluate the influence. This study revealed these communication patterns have long-term effects on the success of OSS projects. In addition, higher centrality within a team promotes communication and exchange of information whereas high project density has negative effects on communication, thus a balance between density and centrality measures is required to ensure long-term success in OSS projects. 

Gerosa et al.~\cite{Gerosa2021} studied the factors affecting the motivations of contributors that drive them to contribute to OSS projects using a survey and compared the findings stated in previous work by Hars et al.~\cite{Hars2001}. The survey was conducted in 2020 and revealed that OSS contributors tend to contribute more because of intrinsic reasons. Also, some motivations have not been shifted since the last survey, however, social aspects such as altruism, kinship, and reputation have drastically increased. Moreover, older contributors are more interested to help (altruism) and tend to increase social interaction while young developers mostly contribute to OSS projects to improve their resumes.

As opposed to the previous studies, our study considers the communication channel of the comments features provided by the JIRA platform in order to extract the knowledge in the form of concepts for each developer.

\subsection{Representation of domain expertise}

Extraction of concepts from source code has already been studied in the literature. Abebe et al.~\cite{Abebe2010} employed natural language processing (NLP) to extract ontologies from program code. The extracted ontologies turned out to be helpful in reducing the search space when programmers query for code location using ontology concepts.

Dey et al.~\cite{Dey2021} proposed \textit{Skill Space} to conceptualize domain expertise. The proposed method can be applied to individual developers, projects, programming languages, and APIs. The aim of the study was to analyze the assumptions regarding whether developers focus on similar new APIs, projects, or languages they frequently use. One of the major contributions of this study is the ability to compare the three entities (developers, projects, and APIs) in the same space, thus enabling developers to evaluate the expertise of a new developer aspiring to collaborate. However, the study is limited to the APIs for the capturing of domain expertise and did not consider other sources such as communication and collaboration within teams.

Omoronyia et al.~\cite{Omoronyia2009} proposed a contextual awareness model based on developers' activities on source code. The model illustrates the association of developers and artifacts with the work context, such as which tasks or artifacts consumed the highest effort among the developers, and further provides a social graph that highlights the potential bottlenecks of modifying or removing tasks or artifacts.

Cetin et al.~\cite{Cetin2020} categorized developers in a software development team into three categories: Jacks, Mavens, and Connectors. Algorithms were proposed for each category using artifact traceability graphs. The study was evaluated on three OSS projects by using the extraction of top commenters on these artifacts in order to validate the results of the model implementing these algorithms. Results showed the proposed model successfully identified the individuals belonging to the aforementioned categories in the projects.

Vadlamani et al.~\cite{Vadlamani2020} studied the developer's expertise based on the findings from Github\footnote{\url{https://github.com/}} and StackOverflow\footnote{\url{https://stackoverflow.com}} collaborative platforms. An exploratory survey was conducted with 73 developers as subjects to understand their perspectives on contributing to collaborative platforms. The results from the quantitative analysis revealed that knowledge and experience are the most suitable predictors of expertise. However, the results from the qualitative analysis show that the soft skills of the developers are of core importance in determining expertise. The study concluded that an individual should possess both in order to be an expert.

In contrast to the aforementioned studies, rather than focusing on the solution space, such as source code and APIs, to determine the domain expertise of developers, the underlying focus of this study is to provide a coarser-grained and practical representation of domain knowledge through problem space such as the emergence of issues and the issue focused involvement of developers.

\subsection{Topic modeling in software engineering}

Topic modeling has been utilized in the past to classify documents into various topics for the purpose of sentiment analysis~\cite{Lu2011,Xianghua2013}, detecting toxicity in text~\cite{Obadimu2019}, or generating recommendations~\cite{Jiang2017}. For instance, Hong et al.~\cite{Hong2010} utilized topic models to predict potential categories of Twitter messages. Two models were evaluated in this study: Latent Dirichlet Allocation (LDA) and Author-Topic model. The results reveal that the Author-topic model performs significantly worse than the standard LDA model.

Jiang et al.~\cite{Jiang2017} proposed approaches to recommend the most suitable commenter for a pull request. These approaches were evaluated using 8 projects as subjects in the case study. Results of this study show that the activeness-based approach outperforms the rest of the approaches including text and file similarity. This study also suggests that the activeness attribute is of core importance in order to recommend suitable commenters.

Panichella et al.~\cite{Panichella2013} introduced a novel solution known as LDA-GA to build LDA models tailored specifically to perform software engineering activities. These activities include traceable link recovery, feature location, and labeling of software artifacts. The solution is based on genetic algorithms in order to determine the best LDA configurations tailored to software engineering activities. LDA-GA is compared with existing heuristics for LDA configuration. The results show that LDA-GA is capable of identifying the best suitable configurations, thus improving the accuracy of LDA models employed for software engineering datasets.

In our work, we aim to employ a topic modeling technique in order to extract and classify the topics present within the issues and comments exhibiting developers' interaction, and monitor the changes that occur over the course of the development of the project.

\subsection{Word representations in software engineering}

Word representations are an important way to understand and describe natural language. It has been employed in many software engineering activities.
Trivedi et al.~\cite{Trivedi2020} proposed a deep learning approach based on the LSTM model to detect the existence of nano-patterns in code. To achieve this, the code is first preprocessed by utilizing word embeddings in order to train the model. The approach is evaluated on a Java code resulting in an accuracy of 88.3\% in predicting nano-patterns in the code. Ferrari et al.~\cite{Ferrari2019} used word embeddings to understand the variations in the terms and identify possible ambiguities in the requirements from different domains.

Guo et al.~\cite{Guo2017} introduced a solution based on deep learning in order to detect traceable links between source artifacts and target pairs. The proposed solution employs word embeddings and a customized recurrent neural network (RNN) model in order to generate these traceable links. A large corpus aggregated from PTC and Wikipedia is used to evaluate the approach. The results showed significantly higher MAP scores for the proposed approach as compared to previous tracing methods VSM and LSI.

Ferrari et al.~\cite{Ferrari2019} proposed a natural language-based approach to detect ambiguous terms in requirements from different domains and provided an ambiguity score in order to rank them. The aim is to construct domain-specific language models and compare their word embeddings to understand the variations in the terms and identify possible ambiguities in the requirements from different domains. The results show that in some cases the proposed approach was effective while in most cases evaluation was not accurate due to the presence of high-level abstract terms and other factors.

% Shafiq et al.~\cite{Shafiq2021} proposed a recommendation approach to allocate incoming tasks to the most suitable roles in software development. The approach utilized textual attributes of tasks and converted them into word embeddings, which were later used as features for training an LSTM model. A benchmark study has also been conducted to compare the performance of other alternate models. The results of this study demonstrated \textasciitilde 70\% accuracy by evaluating the LSTM model on 10 software projects implying a significant increase over the rest of the alternatives.

Shafiq et al.~\cite{Shafiq2021a,Shafiq2021} proposed recommendation approaches to prioritize issues and allocate incoming tasks to the most suitable roles in software development. The approaches utilized textual attributes of issues and tasks and converted them into word embeddings, which were later used as features for the training of machine learning models.

In this paper, we are employing TF-IDF~\cite{Ramos2003} representations of the bag of word~\cite{Zhang2010b} embeddings in order to capture the relevance of words within issues and their comments. These representations are further employed in the process to construct the \textit{ConceptRealm}, which is described in detail in Section~\ref{TR}.

\subsection{Turnover in software development}

The developer volatility in organizations is inevitable as developers may switch to other teams within an organization or even join other organizations leading to a turnover. This induces a loss of expertise and a great gap in the knowledge possessed by the individual developer leaving.

A study showed that a newcomer takes a considerable amount of time to get a deep understanding of the project similar to the leaving core developer leading to a great loss in productivity~\cite{Izquierdo-Cortazar2009}. Moreover, the more  orphaned code there is in the project, the more defects it will produce~\cite{Otte2008}.

Robillard et al.~\cite{Robillard2021} studied the loss of knowledge induced by developer turnover by conducting interviews with 27 practitioners and managers from 3 companies. The study pointed out various dimensions of turnover. For instance, leaving developers might be available later for knowledge transfer. On the other hand, results also showed that developers who temporarily leave the organization have the same impact as the permanently leaving member as experts are not available during the desperate times of resolving newly emerging issues.

Bao et al.~\cite{Bao2017} investigated the most effective machine learning model to predict developers that are about to leave. Data were obtained from monthly reports that were submitted by 3638 developers in 6 years. The results of this study also showed the random forest appeared to be the best among other alternatives and revealed the most important factors that influence the turnover of developers.

Rigb et al.~\cite{Rigb2016} assessed the risk of developer turnover and investigated ways to cope with this risk to make projects more flexible. The analysis was performed on two projects: one closed-source project Avaya and one open-source project Chrome.
Mockus et al.~\cite{Mockus2010} studied the impact of core developers leaving the project using Avaya as a subject in a case study.

%% file: motivation.tex
%% INTEGRATED IN OTHER SECTIONS ALREADY!!!!

% \subsection{Motivation}

Changes in human resources are inevitable throughout the development process of software projects. For example, an active developer may not work on the same module for an indefinite period or may leave for other reasons~\cite{Xuan2012}. Therefore, the addition of new developers and changes in their priorities are unavoidable. This will lead to an imbalanced knowledge distribution between the old team members and the new ones. Project managers, therefore, will have the task to strike a balance in task assignment between an experienced and a less experienced developer so that knowledge is sufficiently spread across the team.

In addition, prior studies have shown that the developer turnover rate is relatively higher in OSS projects compared to closed-source projects~\cite{DeMelo2013}. Therefore, many companies call for risk managers to avoid such situations and improve developer retention rate~\cite{McManus2012, Lin2017}. For this purpose, this study aims at constructing a general representation of domain knowledge denoted as the \textit{ConceptRealm} within a team, which would help to identify the concepts possessed by the key developers and how they change over time. Furthermore, this coarse-grained representation of developers' domain knowledge would allow managers to be aware of their team's dynamics and valuable information surrounding the issues emerging within the team for newcomers. %We construct the \textit{ConceptRealm} using a wide spread of projects obtained from Jira repositories further elaborated later in Section~\ref{dataset}.

\subsection{Illustrative example}
\begin{figure}[!htbp]
  \centering
  \includegraphics[width=0.5\linewidth]{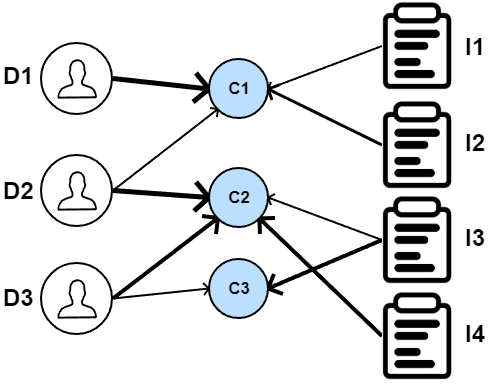}
  \caption{Team and Dev-level concepts association - D (Developers), C (Concepts), I (Issues)}
  \label{CI}
\end{figure}

Consider a simplified development scenario snapshot depicted in Fig.~\ref{CI} comprising three developers D1 to D3, three concepts C1 to C3, and four issues I1 to I4. The arrows among these elements represent the degree to which an issue, a developer, is associated with a particular concept (thicker arrow indicating stronger association). 
Section~\ref{sec:definition} then describes in more detail how we derive these concepts and how we create the relations between these elements.

From Fig.~\ref{CI}, we observe that two developers are knowledgeable in concepts C1 and C2, while only one developer (D3) is familiar with C3. If D3 were to leave the team, no remaining developer would be in a good position to handle new incoming issues associated with concept C3. Whether this is a problem depends on whether C3 is an important concept. In this work, we treat frequency as a measure of how many issues are associated with a concept (again, the formalization of frequency is provided in Section~\ref{EM}).   
To this end, we need to understand how the frequency of concepts changes over time and how that change is reflected by developers' level of familiarity with these concepts.

%% file: definition.tex
\subsection{Concept}
We define a concept as a set of highly coherent terms that appear in the text of an issue. A text may belong to either title, description of the issue, or the comment made by the author on that issue. Each term has a probability indicating how strongly the term is associated to the concept. The total number of terms in each concept may vary from one project to another, however, the terms with the highest probabilities (top 10) belonging to a concept are considered in this study.

These highly coherent terms are essential to describe the tacit knowledge of a particular issue or a comment.
Abebe et al.~\cite{Abebe2015} describe a concept as a domain (e.g., software development, project configuration) or an implementation (e.g., data structures such as arrays and lists) knowledge.
In this paper, we focus on issues and comments for extracting concepts, hence our concepts describe primarily the problem domain and to a lesser extent the solution/technical domain.
There is a designated weight assigned to each association of issues/comments to the concept. This weight is used to indicate the strength, i.e., how closely aligned the concept is to the respective issue or comment. 

An example of how concepts along with their associated weights are obtained from an issue is shown in Fig.~\ref{PracticalExample}. This figure further highlights that the sum of weights ($w_{\mathit{i}}$) for all concepts describing a single issue ``I1'' equals 1. Similarly, the sum of weights ($w_{\mathit{c}}$) for all concepts describing a single comment made by the developer ``D1'' equates to 1. In essence, there are two major elements required for the construction of the \textit{ConceptRealm}: (1) issue-level concepts and (2) comment-level concepts.

\subsubsection{Issue-level representation}
The issue-level concept representation refers to the concepts appearing in the issues. We derive these concepts from the title and description of the issues. Each concept associated with an issue represents the domain knowledge regarding that particular issue.

\subsubsection{Comment-level representation}
The comment-level representation, on the other hand, refers to the concepts appearing among the developers, mainly through the source of comments. We derive these concepts using the comment's body. This level would help us understand invaluable insights, such as the identification of key developers in the team (the concept keepers), distribution of concepts within developers, and ultimately help us monitor the impact on change in these concepts when keepers leave.

\begin{figure}[!htbp]
  \centering
  \includegraphics[width=1.0\linewidth]{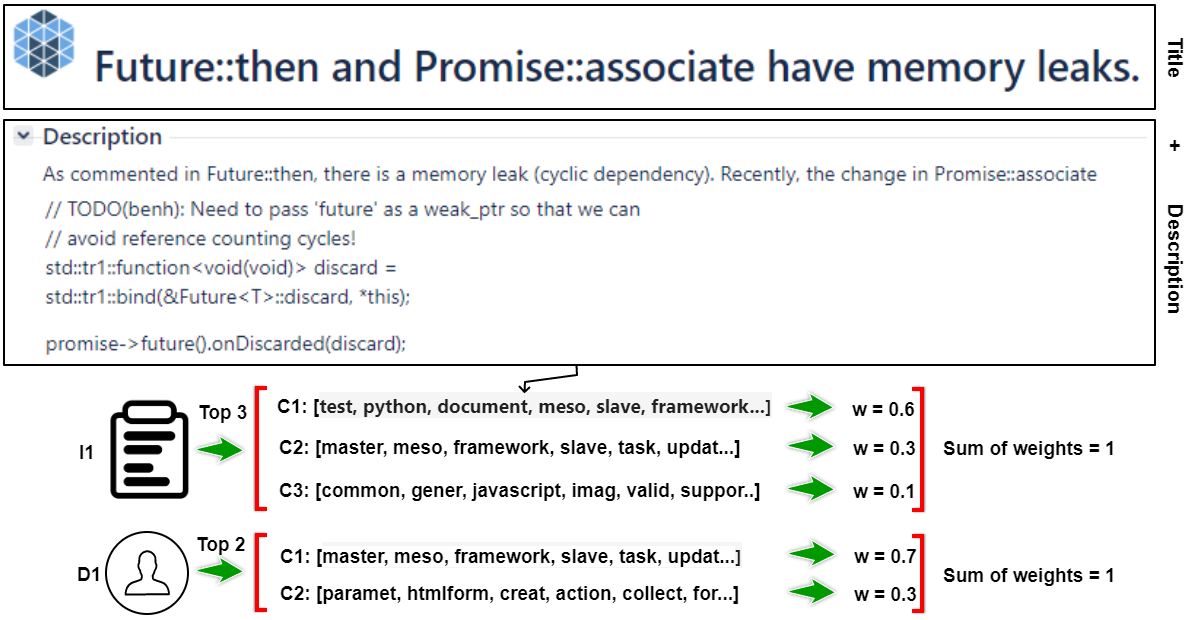}
  \caption{Representation of concepts - I (Issue), C (Concepts), D (Developer), and W (Weight)}
  \label{PracticalExample}
\end{figure}

\subsection{Evaluation metrics}\label{EM}

%We evaluated the \textit{ConceptRealm} by testing our postulated hypothesis within the posed research questions of this study. 

% INTRODUCE frequency
To gauge the rate at which the concepts appear over the years within the team and among the developers, we introduce ``frequency'' as an indicative metric. As opposed to weight, the frequency metric refers to the popularity of a concept showing an approximation of how much a developer or team focuses on the concept with respect to time windows. This metric allows us to quantify each concept with respect to each time window and further compare across multiple projects.

\begin{figure}[!htbp]
  \centering
  \includegraphics[width=0.8\linewidth]{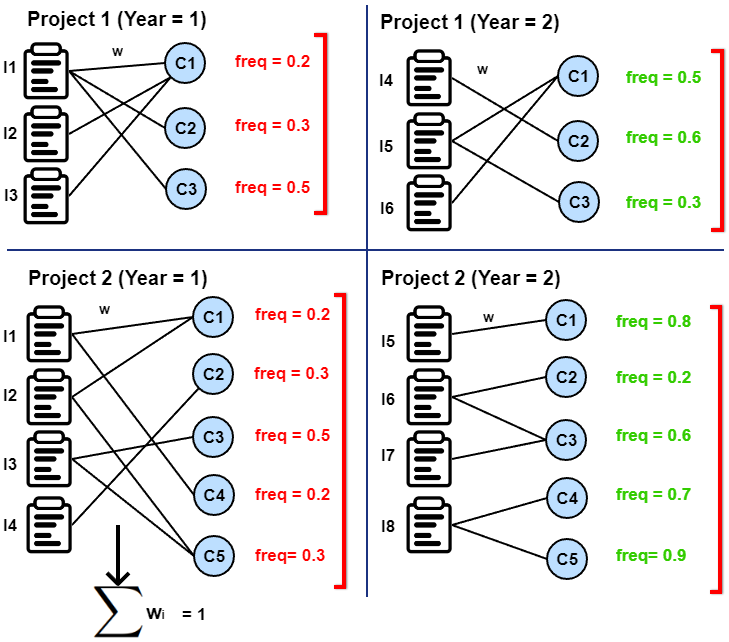}
  \caption{Change in concept frequency}
  \label{relativeImp}
\end{figure}

As the number of concepts varies across the teams we need an additional processing step to ensure the concept frequency changes can be compared across teams.
For example, in a team with five concepts, a single concept might more easily experience a frequency increase of 0.1, compared to a team with 20 concepts, as an increase in the frequency of one concept always comes with the decrease of frequency of other concepts (recall Fig.~\ref{PracticalExample}) and vice versa. 
Hence, directly using frequency will skew teams with fewer concepts to exhibit higher frequency variation than teams with many concepts.   
To this end, we scale the ``frequency'' metric with the number of concepts. 
%We also introduced a metric known as relative ``frequency'' indicating the rate of concepts which appeared over the years. 
Fig.~\ref{relativeImp} shows the effect of number of concepts on the frequency of the concept. %The reason of considering a relative measure for evaluation is to ensure that if a project has a very low number of concepts, its frequency would be negligible as opposed to the project with greater number of concepts. 
To analyze teams and individual developers, ultimately we introduce two frequency metrics.
For the issue-level representation, the relative frequency of a concept is measured using the following equation:

\begin{equation}
\label{eqteam}
% \begin{align}
\textit{$freq_{issue}(C)$} = \frac{\sum_{i=1}^{n}w_{\mathit{i}}}{n} * |C|
% \end{align}
\end{equation}

where $w_{\mathit{i}}$ is the weight value for each concept per issue $i$, $n$ is the total number of issues, and $C$ is the set of concepts in the project. 

For the developer-level representation, the frequency is measured as follows:

\begin{equation}
\label{eqdev}
% \begin{align}
\textit{$freq_{comment}(d,C)$} = \frac{\sum_{i=1}^{n}w_{\mathit{c}}(d)}{n} * |C|
\end{equation}

where $w_{\mathit{c}}(d)$ is the weight value for each concept per comment $c$ for a developer $d$; $n$ is the total number of comments made by $d$, and $C$ is the set of concepts in the project. From this follows that the sum of concept frequency values at the team level as well as the sum of concept frequency values summed across all developers each equal $|C|$.

%% file: studydesign.tex
% INTRO STUDY  DESIGN with FIG 4
%\subsection{Study design}
Overcoming the aforementioned problems due to turnovers during a project requires a careful understanding of the knowledge in possession of the individuals within a team. The goal of this study is to provide a representation constituting this knowledge called the \textit{ConceptRealm}, ultimately providing invaluable insights to the managers.
The guidelines to direct this research are provided by Basili et al. and Runeson et al.~\cite{Basili,Runeson2009}. 

The \textit{Purpose} of this study is to investigate the predictive ability of these concepts and to monitor the change in these concepts during the course of project evolution.
The \textit{Issue} is the imbalanced distribution of knowledge caused by the change in the team structure. The aim is to monitor the transitions of concepts emerging at both the team and the developer level.
The \textit{Viewpoint} is of the managers or team leads that can leverage the \textit{ConceptRealm} to identify whether such concepts align with the developers and predict the impact of potential leaving members in the team.
The \textit{Process/Context} of this study encompasses the issues and comments made by the developers obtained from OSS projects in issue tracking platforms. %, the emerging concepts within these issues and comments, and weighing their association using frequency as a metric.

% THEN REQS 
\subsection{Research questions}
In the remainder of this paper, we focus on the following six research questions.
% Research questions help guide the research and act as a systematic mean to reach the goal. To this end, we have formulated the following research questions in order to address the aforementioned goals.

%\begin{enumerate}[font={\bfseries},label={RQ-\arabic*:},leftmargin=3.1\parindent]
% \item What sort of concepts emerge within a software development team?

% \textbf{Rationale:}
% This question refers to the type of concepts extracted from textual features present in the issues titles, descriptions, and comments. Whether the concepts are more related to the implementation or the domain in general. A concept may represent a domain (e.g, software development, project configuration) or an implementation (e.g, data structures such as arrays, lists) knowledge~\cite{Abebe2015}.

%\item 
\textbf{RQ1: Are the extracted concepts meaningful?}\\
% SRQ-1: is it common to have stable concepts or are they changing?
%\textbf{Rationale:}
In order to validate that the extracted concepts are indeed meaningful and not just noise, we investigate whether concepts help to predict who will work on an issue. To this end, we assume that a team member familiar with a concept in the past is more likely to work on a future issue (related to that concept) compared to the most active member of the team. %that i  s with strongest concepts in the past later gets assigned to the new issues which are classified as these concepts. Thus, we compare developers associated to concepts in the past with the developers associated to issues with similar concepts in the future vs random developers associated to concepts.

\textbf{RQ2: To what degree do concepts change over time at the issue-level and the comment-level?}\\
%\textbf{Rationale:}
This question aims to provide insights into the typical extent to which concepts within a software development team evolve over the course of multiple years, ultimately highlighting the changes in the knowledge distribution within the team. To this end, we observe whether the changes in concept frequency at the issue-level are also reflected at the comment-level and thus whether the issue-level concept changes are representative of an individual developer's interest in concepts. %Comparing an individual developer's interest in a concept to overall concept frequency allows us to measure the distribution of concept familiarity in a team and hence detect concept keepers with potentially significant impact upon leaving the team.
The point to observe is whether concepts remain stable, and if not, how much does the change of concepts differ for different teams?

%\item How does concepts frequency changes if a developer leaves?
% SRQ-1: is it common to have stable concepts or are they changing?
\textbf{RQ3: Are concept metrics able to measure the effect of a leaving developer on team knowledge?} \\
%\textbf{Rationale:}
In order to support a manager or team lead in estimating/predicting the impact of a leaving member, we need to show that developers that act as a  keeper of a concept indeed may result in the remaining team members becoming less able to contribute knowledge associated with this concept. We hypothesize that the more a keeper holds the knowledge of a concept, the more that concept will drop in significance upon that keeper's departure. Additionally, we hypothesize that a developer that shares the weakest level of concept familiarity as other team members will have a negligible effect on the raise for drop in the concept's frequency upon leaving.

\textbf{RQ4: Is there a difference in project/concept distribution for those developers in the negative quadrant compared to the positive quadrant}

% -- Look for projects with data points in the positive quadrant have also data points in the negative quadrant???

% -- Look whether devs strongest concept aligns with the highest issue-level concept???

We aim to investigate the difference in the concept distribution for developers that reside in the negative quadrant, i.e., their concept decreased when they became inactive, and in the positive quadrant, i.e., the concept increased on their departure from the project. We also observe these differences within projects to have a general perspective of the distribution of concepts.
% This questions refers to the analysis of the distribution of concepts among projects.

% How much can we improve developer recommendation based on the concept distribution????
\textbf{RQ5: What are implications for assignee recommendation algorithm that should also result in more evenly distributed knowledge?}

% --two clear datasets (equal/unequal)

% --cross test

In this question, we want to understand how \textit{ConceptRealm} can help improve the current assignee recommendation algorithms in achieving a balanced distribution of concepts thus helping in maintaining a similar distribution of knowledge among developers in the project.

\textbf{RQ6: How effective \textit{ConceptRealm} can be in the detection of concepts and identifying the impact of leaving developers in the industry?}

Analysis of our OSS projects leaves an open-ended question of whether the information obtained from \textit{ConceptRealm} can be effective for managers/developers. To investigate further, we evaluated the usefulness of concepts and the effect of leaving members using an industrial case study from Dynatrace.

%\end{enumerate}

\begin{figure*}[!htbp]
  \centering
  \includegraphics[width=1.0\linewidth]{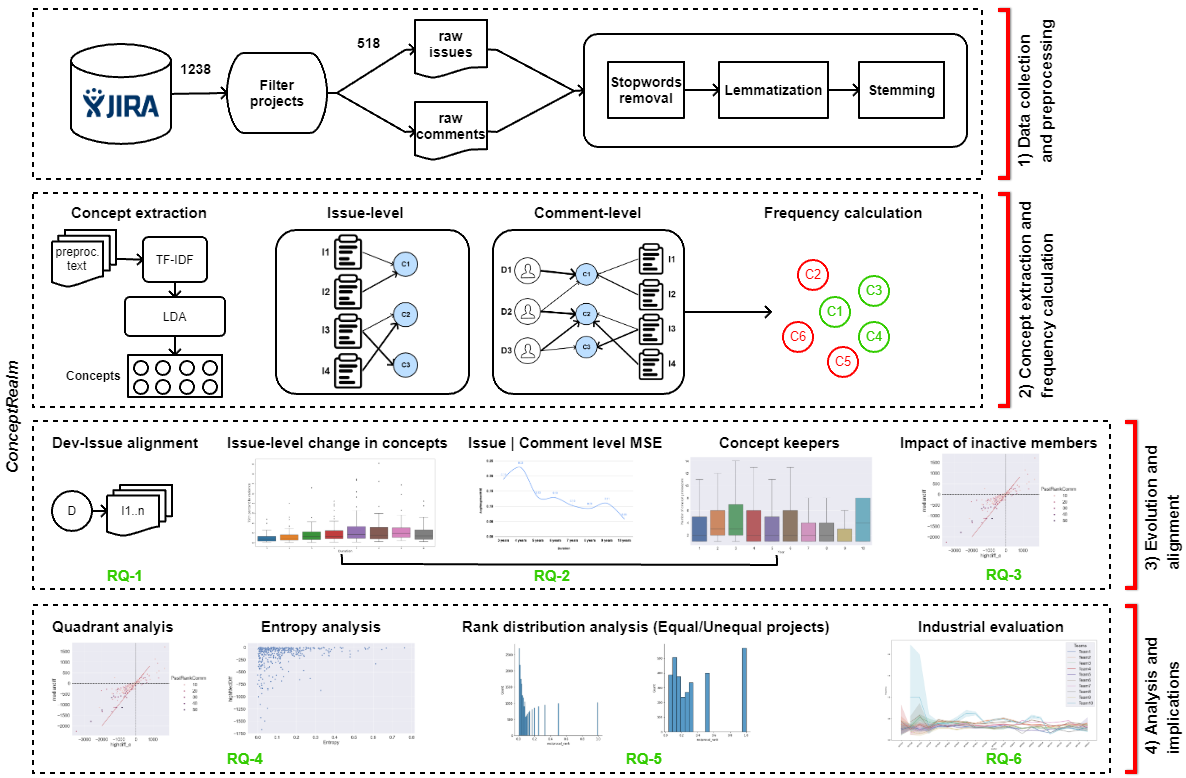}
  \caption{Study design overview}
  \label{RM}
\end{figure*}

% --------------------- TO INTEGRATE THE TEXT ABOVE AND BELOW ----------------
To answer these research questions, we follow the approach depicted in Fig.~\ref{RM}. 
We build on an existing, vetted dataset by Ortu et al.~\cite{Ortu2015a} consisting of Jira issues and comments as briefly introduced further below.
After data preprocessing, we apply concept extraction and subsequently concept frequency metrics at the issue and comment-level. This data is then subject to further analysis to answer RQ1 to RQ6.
%The overall goal of this study is to understand how concepts (knowledge) fluctuate in a team over time and are distributed among developers. To this end, we utilize data collected from the well-known issue tracking tool Jira and identify the concepts represented by various issues handled by development teams. The method comprises the construction of the OSS dataset, which essentially involves preprocessing of the data, and extraction of concepts from the data. Lastly, we evaluate our research questions based on the analysis of the data. Fig.~\ref{RM} refers to the method followed through this study. This study is divided into three phases and the elements of each phase are discussed below in detail.

\subsection{Dataset}\label{dataset}
The base dataset provided by Ortu et al.~\cite{Ortu2015a} contains issues from 1238 projects from four Jira repositories, which include Apache\footnote{\url{https://issues.apache.org/}}, Spring\footnote{\url{https://jira.spring.io/}}, JBoss\footnote{\url{https://www.jboss.org/}}, and CodeHaus.\footnote{\url{https://www.codehaus.org/}} Note that the projects in the dataset are restricted to 2015. However, this would not impact the results of this study, as the study~\cite{Gerosa2021} showed that aside from social aspects, motivations to contribute to OSS projects among developers have not shifted since the commercialization of OSS development. Presently, older contributors may lead to better knowledge distribution but it does not hold for every project. Apart from this, we did not find any significant change in developers' OSS interactions till this study was conducted~\cite{Gerosa2021}.

Most vital for our analysis, they have ensured that all comments are from actual developers and not from infrastructure bots, such as integration servers and build pipelines, which they kept in a separate database table.
We further filtered the number of projects down to 518 to include only projects that have a set of textual terms sufficiently diverse but also frequent enough to build a reliable set of concepts.
Specifically, we applied the following filter criteria:

excluded projects by filtering out most rare (tokens that are present in less than 15 issues (no\_below = 15)) and frequent words (tokens that are present in more than half (no\_above = 0.5) of the project) in the project for the purpose of keeping only the words, which help in capturing the context.

%We extended the dataset provided by Ortu et al.~\cite{Ortu2015a} with the addition of concepts associated with each issue as well as the developer. 

We hypothesized that project age might influence the number of concept changes. Hence we checked the age of projects and clustered them into age groups. In doing so, we found that projects with an age of fewer than 3 years and greater than 10 years were small in number. Thus, to maintain a similar distribution of projects, we grouped the projects into eight age brackets ranging from 3 to 10 years. The final pool of projects is 518 with over 300k issues and 1.3M comments. Table~\ref{DistProjects} provides additional details on how these numbers are distributed per age (year) bracket. Ultimately, we extended this base data subset with concepts and their association with developers and teams for each year. This extended dataset along with the scripts to construct the \textit{ConceptRealm} and to reproduce all results of this paper are available as a replication package~\href{https://doi.org/10.5281/zenodo.5167218}~\cite{anonymous2022}.

\begin{table}[htbp]
\centering
\begin{tabular}{rrrrrr}
\toprule
Age &  Projects &  Issues &  Comments & Devs & Median(Devs) \\
\midrule
3 &        53 &   19556 &  61332   & 2663 & 20.0  \\
4 &        73 &   20310 &  60706   & 3149 &  26.0  \\
5 &        65 &   29539 &  111919  & 5482 & 34.0  \\
6 &        72 &   38946 &  157838  & 6511 & 52.0  \\ 
7 &        73 &   77341 &  308092  & 11052  &  59.0  \\
8 &        57 &   44601 &  203544  & 8548 & 99.5  \\
9 &        63 &   34853 &  140549  & 7839  &  92.5 \\
10 &       62 &   73814 &  257657  & 16326 & 143.0  \\
\bottomrule
\end{tabular}
\caption{Distribution of projects across age brackets}
\label{DistProjects}
\end{table}
% \vspace{-8mm}

\subsection{Data pre-processing}\label{dp}

In order to gain meaningful insights from LDA, the dataset is required to go through a systematic cleaning process. We first eliminated stopwords from the textual features in the dataset, which include the title, description, and comment body of each issue. We further performed the lemmatization process using WordNet~\cite{Miller1995}. Lemmatization refers to the process of extracting the dictionary form of a word -- also known as \textit{lemma} -- while removing inflectional post-fixes. We further performed Porter's stemming process~\cite{Porter1980} on the \textit{lemma}, which refers to the removal of word endings to a stem that could possibly be a word not present in the dictionary.

\paragraph{TF-IDF representation}\label{TR}

The preprocessed issues and comments are then converted to vector embeddings using the doc2bow algorithm. Doc2bow is an implementation provided by the Genism\footnote{\url{https://radimrehurek.com/gensim/}} library to generate bag of words embeddings from the documents (referred to as text of issues and comments in this study). For each word in a document, these embeddings are represented as one-hot encoded vectors. 
% An example illustrating this process of conversion is shown in Fig.~\ref{bowRep}. 
These word embeddings are later converted to Term Frequency-Inverse Document Frequency (TF-IDF) vector space. TF refers to the number of times a word has appeared in an issue or a comment whereas, TF-IDF~\cite{Ramos2003} is a simple and efficient algorithm to match words that are relevant to the issue based on the entire corpus. This TF-IDF representation is then used as input to the construction of the LDA model.

% \begin{figure}
%   \centering
%   \includegraphics[width=1.0\linewidth]{bowRep.png}
%   \caption{Conversion of sentences to bag of words embeddings}
%   \label{bowRep}
% \end{figure}

\subsection{Concept extraction}\label{CE}

In order to obtain a meaningful set of terms, we have trained an LDA model on the aforementioned TF-IDF representation of issues and comments data for each project. LDA~\cite{Blei2003} is a statistical model commonly employed to generate and classify document topics. We are referring to these topics as concepts throughout this paper. The LDA model is then used to generate the concepts that best capture the spectrum of issues. The LDA model is further used to associate each issue with the respective concept. In essence, once the LDA model is trained on the corpus we use the model to generate probability scores for each concept given an issue. These probability scores indicate how close the concept is to the issue. Similarly, with the comments on each issue made by the developers, we have used the same LDA model to associate these comments with the respective concept. We call this resulting association the \textit{ConceptRealm}. %, which provides an ability to inspect the issue concepts and whether the developers gain these concepts when applied to time intervals. 
%  Fig~\ref{ConceptRealm} shows the internal steps undertaken in the \textit{ConceptRealm}. This representation will further help to investigate important questions such as does the same concept emerges in later years during the project or whether the concept depreciates if the associated author leaves the project.

% \begin{figure}
%   \centering
%   \includegraphics[width=1.0\linewidth]{ConceptRealm.png}
%   \caption{ConceptRealm}
%   \label{ConceptRealm}
% \end{figure}

\paragraph{Sanity check}\label{SC}

\begin{figure}[!htbp]
  \centering
  \includegraphics[width=0.8\linewidth]{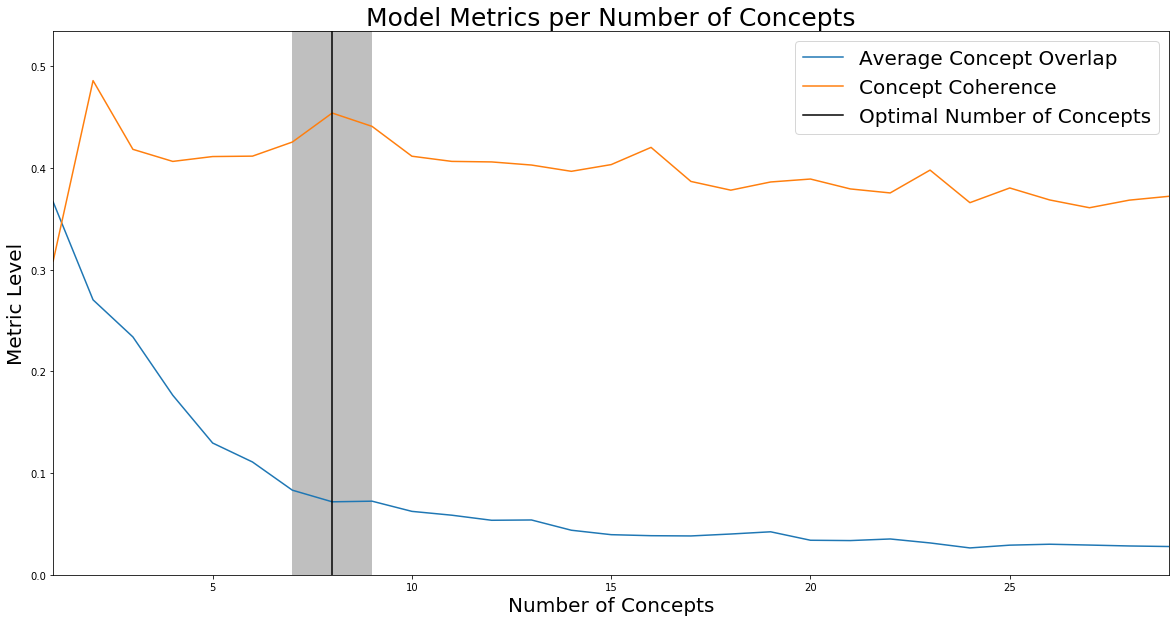}
  \caption{Example of determining the optimal concept number}
  \label{idealConceptNumber}
\end{figure}

We have followed the guidelines described by Panichella et al.~\cite{Panichella2013} in order to get the optimal LDA configuration for each project. We first created LDA models fed with the same text corpus and a different number of concepts ranging from 1-30. We chose this range as the average optimal number of concepts obtained for each project later appeared to be less than 20. Also, there is a low risk of overestimating the number of concepts as compared to underestimating the number of concepts as suggested by Wallach et al.~\cite{Wallach2009}. For every LDA model created, we then calculated the Jaccard similarity value (as suggested by Abebe et al.~\cite{Abebe2015} and exemplified in equation~\ref{eqJac}) for each concept and compared it with all concepts (Concept Overlap). We also calculated the coherence (as expressed in equation~\ref{eqCoh}) within all the concepts across the LDA models. The coherence was calculated using the best performing measure ``Cv'' as supported by the benchmark study~\cite{Roder2015}. We used the coherence module from Gensim\footnote{\url{https://pypi.org/project/gensim/}} library, which is well known in concept modeling and NLP. Finally, we selected the optimal (i.e., highest coherence and lowest concept overlap) number of concepts to build the final LDA model for each project. We calculated this optimal number by taking the maximum from the difference between coherence and overlap, i.e., $MAX(coherence - overlap)$. An example of determining the optimal number of concepts can be seen in Fig.~\ref{idealConceptNumber}. The horizontal lines represent the average concept overlap and concept coherence, whereas the vertical line identifies the optimal number of concepts for the given project.

\begin{equation}
\label{eqJac}
% \begin{align}
\textit{jaccard($C_1,C_2$)} = \frac{|C_1 \cap C_2|}{|C_1 \cup C_2|}
% \end{align}
\end{equation}

\begin{equation}
\label{eqCoh}
% \begin{align}
\textit{coherence} = \sum_{i<j}{ \phi (w_i,w_j)}
% \end{align}
\end{equation}

%% file: results.tex
%In order to perform quantitative analysis, we have employed most commonly used data manipulation libraries for Python including: Pandas, and Numpy.
In this section, we answer the above-introduced research questions.%, with discussion and implications of our findings found in the subsequent Section~\ref{sec:discussion}.

\subsection{Validating the extracted concepts}
Addressing RQ-1, we analyzed the meaningfulness of concepts by comparing the past concepts associated with developers and whether they are assigned to new issues exhibiting similar concepts vs most active developers of the same project. 
Note that the goal is not how well we can predict the assigned team member, but whether our approach produces concepts that are meaningful enough to describe a team member's familiarity with the domain and hence being likely to work on such an issue.

We first selected a random year from each project and divided it into half based on the issue creation date. The reason for selecting one random year is due to the fact that we want to avoid the evolution of these concepts, which would be significant when evaluating for more than a year. Note that these concepts are different than the year-wise concepts generated by the LDA model to address the aforementioned research questions. Consequently, the past data (i.e., issues from the year's Q1 and Q2)  become the training set and future data (i.e., issues from the year's Q3 and Q4) become the test set. We then trained the LDA model on the training data and generated a set of concepts. Later we used this model to classify each issue and comment in the data as one of the dominant concepts.
For each issue in the test set (second half of a year) we determine (a) how well the assigned developer matches the issue's strongest concept (i.e., via the frequency value) and (b) how well the most active developer matches the issue's strongest concept. Then we derive the mean over these two
'groups' and determine the difference. These two groups (lists of frequency values) as also used as the input to the pairwise t-test. (The diff itself is not used for the t-tests).
Consequently, we performed a pairwise t-test to measure if there is a significant difference among the developers who become assignees for the new issues exhibiting similar past concepts vs most active developers from the test sample.
Before applying, we ensured that the assumptions of the test are met, i.e., independent samples from the same group and normally distributed data. This test was applied to 243 projects\footnote{For this RQ, concepts are extracted from a smaller data set, i.e., half a year of comments and issues, hence some projects didn't yield sufficiently frequent words as outlined in the section \ref{dataset} and thus were not included for answering RQ1.}. % as the rest of the projects were excluded using the same filter as described in Section~\ref{dataset}.
For the pairwise t-test, two groups are being compared for each project, (1) set of tuples: developers with strongest concepts assigned to issues and (2) group of most active developers. The test gives as output the means for both groups, a higher mean for a group indicates stronger alignment with the issues which, in our case, is the former group. The accuracy metric in the table~\ref{meandiff} is just added to show how many developers are actually assigned to the issues that had the strongest similar concept in the past.

The results of the t-test showed that out 211 of 243 projects demonstrate a significant difference (p-value~\textless~0.05) supporting the hypothesis that developers highly associated with a concept in the past are more likely to be assigned to new issues that are also aligned with the same concept compared to a most active team member. Table~\ref{meandiff} shows the average difference in means between the variables along with the number of projects (p-value~\textless~0.05) with respect to each project duration range. The table also shows accuracy which indicates the percentage of developers that were assigned to issues aligning with their strongest concepts. The high number of accuracy values for projects across all age brackets further strengthens our assumption.
Note that with this experiment we don't suggest a new assignee recommendation metric but demonstrate the usefulness of determining concepts, hence that the extracted concepts are neither arbitrary nor noise.

% As compared to the OSS projects, we believe that closed-source project should exhibit similar results regardless of the differences.

\begin{table}
\centering
\begin{tabular}{ccccc}
\toprule
\multirow{2}{*}{Duration}  & \multirow{2}{*}{Projects} & \multirow{2}{*}{totalProjects} 
&  \multirow{2}{*}{meanDiff} & \multirow{2}{*}{meanAccuracy} \\[3pt]
(yrs) & (P < 0.05) & & & (\%)
\\
\midrule

% %Random devs
% 3 &          22 &            25 &      0.68 & 76.99 \\
% 4 &          24 &            31 &      0.72 & 81.48 \\
% 5 &          24 &            33 &      0.48 &  80.11\\
% 6 &          24 &            30 &      0.73 & 81.30 \\
% 7 &          37 &            40 &      0.56 &  85.56\\
% 8 &          18 &            24 &      0.56 & 83.72 \\
% 9 &          13 &            21 &      0.51 & 82.39 \\
% 10 &          25 &            34 &      0.43 &  79.60\\

% Most Active dev
3 &          18 &            22 &      0.67 &     75.52 \\
4 &          27 &            34 &      0.58 &     82.13 \\
5 &          29 &            32 &      0.88 &     78.72 \\
6 &          31 &            34 &      0.78 &     79.14 \\
7 &          34 &            39 &      0.72 &     82.21 \\
8 &          20 &            22 &      0.94 &     83.68 \\
9 &          22 &            24 &      0.67 &     75.75 \\
10 &          30 &            36 &      0.68 &     81.13 \\

%  \rowcolor{green}
\bottomrule
\end{tabular}
\caption{Pairwise t-test results: average (across projects) of mean difference (within a project) in comment-level concept frequency between assigned developers vs most active developers}
\label{meandiff}
\end{table}

\begin{center}
    \fbox{
	\begin{minipage}[c]{0.95\linewidth}
		\textbf{Summary of RQ-1:} Extracted concepts are meaningful as for a large majority of projects (211/243) they allow to better predict the issue's assignee than choosing the most active developer in the project. 
% 		It allow us to predict the likelihood of developers assignment to new issues i.e., developers previously associated to the concepts are later associated to same concepts emerging in the new issues and comments. This indicates that concepts are indeed useful in understanding the evolution of domain knowledge possessed by the team as well as the developers.
	\end{minipage}
    }
\end{center}

% \subsection{Expert opinion}
% We also validated these concepts by conducting series of informal interviews with the team lead of a well-known organization. The representative agreed to the usefulness of the concepts and stated that these concepts are indeed representative of the developer's knowledge.

% \subsection{Lessons learned}
% The expert also identified few concepts to be associated with issues that are redundant and automatically generated during a period when they were testing new tools in their team. We believe such noise in the data could result in the generation of concepts that are of minimal or negligible value. Thus, the only way to identify these concepts is to have them checked by a project stakeholder before conducting a deep analysis.

\subsection{Measuring concept evolution}
\begin{figure}[!htbp]
  \centering
  \includegraphics[width=0.8\linewidth]{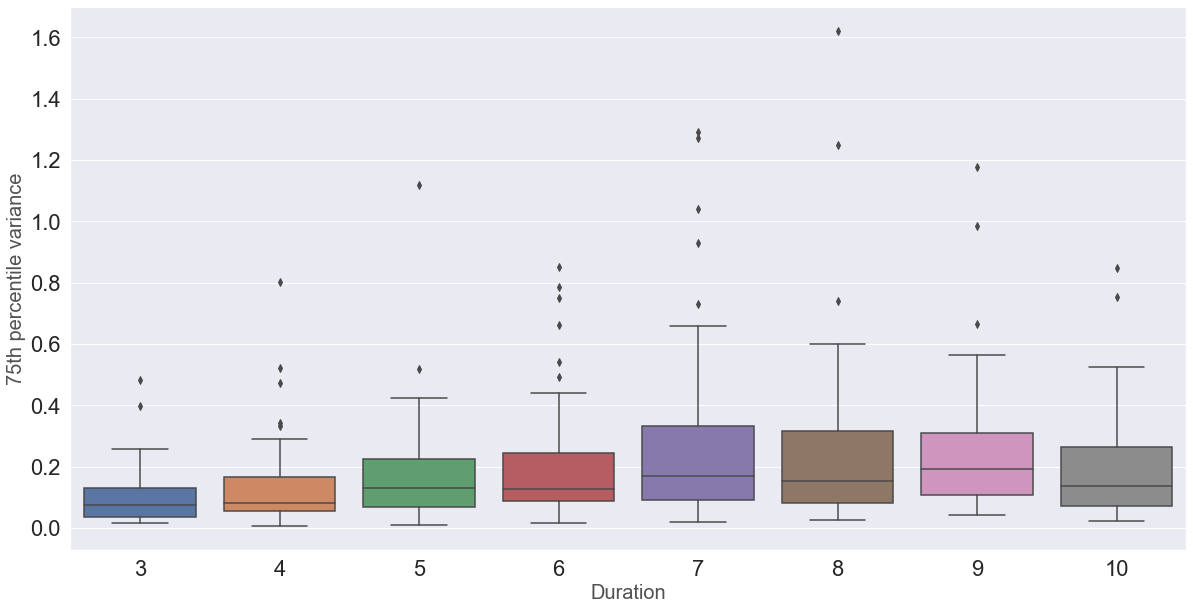}
  \caption{75th-percentile variance of issue-level concept frequency by project duration}
  \label{meanvar}
\end{figure}

Addressing RQ-2: ``To what degree do concepts change over time at the issue-level and the comment-level?'' We calculated the variance of the year-to-year frequency changes for each concept within a project. 
As a single large frequency change of one concept and otherwise stable concepts will result in a larger variance for a 3-year project compared to the same single frequency change in a 10-year project (more data points over which to aggregate) we compare the frequency changes only for projects of the same age. These age groups range from 3 - 10 years. Typically, only a small set of concepts experience a frequency change from one year to the next. Hence, we take the 75th-percentile of concept frequency change variance per project to obtain more insights into how much those more fluctuating concepts change.
The set of 75th-percentile variance values from each project (grouped by age) then produces the boxplots in Fig.~\ref{meanvar}. 
%We reported the 75th-percentile variance of the frequency of concepts. Fig~\ref{meanvar} shows the mean variances of concepts frequency across the groups. 

\begin{figure*}
  \begin{subfigure}{8cm}
    % \centering
    \includegraphics[width=8cm]{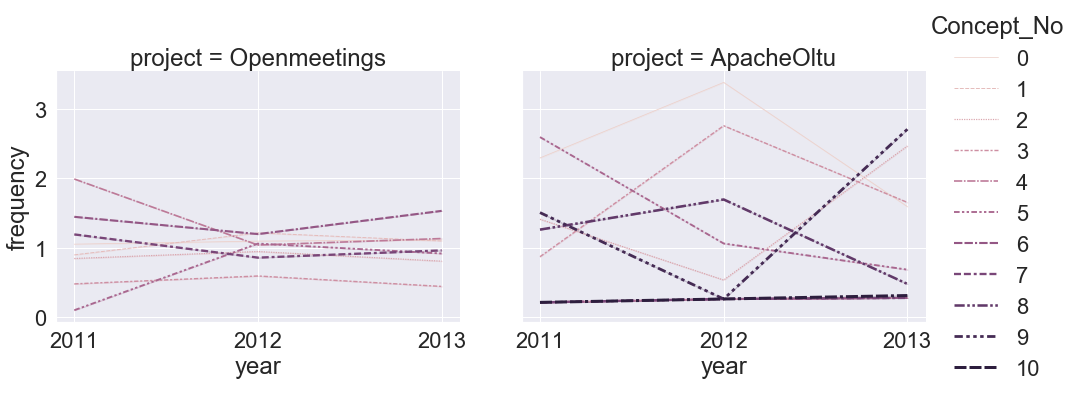}
    \caption{project duration = 3 years}
  \end{subfigure}
 
%   \begin{subfigure}{6cm}
%     \centering\includegraphics[width=6cm]{4yearMinMaxVarProjects.png}
%     \caption{project duration = 4 year}
%   \end{subfigure}
%   \begin{subfigure}{6cm}
%     \centering\includegraphics[width=6cm]{5yearMinMaxVarProjects.png}
%     \caption{project duration = 5 year}
%   \end{subfigure}
  \begin{subfigure}{8cm}
    % \centering
    \includegraphics[width=8cm]{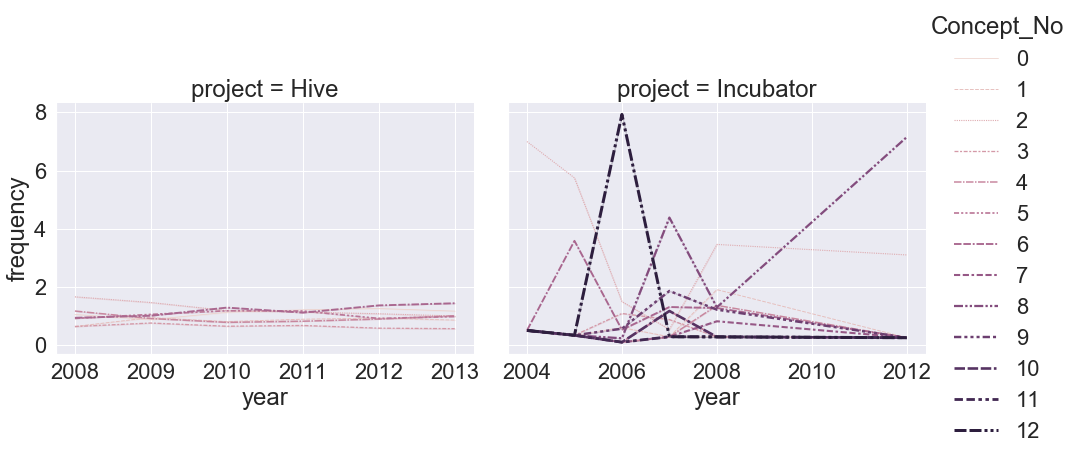}
    \caption{project duration = 6 years}
  \end{subfigure}
%   \begin{subfigure}{6cm}
%     \centering\includegraphics[width=6cm]{7yearMinMaxVarProjects.png}
%     \caption{project duration = 7 year}
%   \end{subfigure}
%   \begin{subfigure}{6cm}
%     \centering\includegraphics[width=6cm]{8yearMinMaxVarProjects.png}
%     \caption{project duration = 8 year}
%   \end{subfigure}
%   \begin{subfigure}{6cm}
%     \centering\includegraphics[width=6cm]{9yearMinMaxVarProjects.png}
%     \caption{project duration = 9 year}
%   \end{subfigure}
  \begin{subfigure}{8cm}
    % \centering
    \includegraphics[width=8cm]{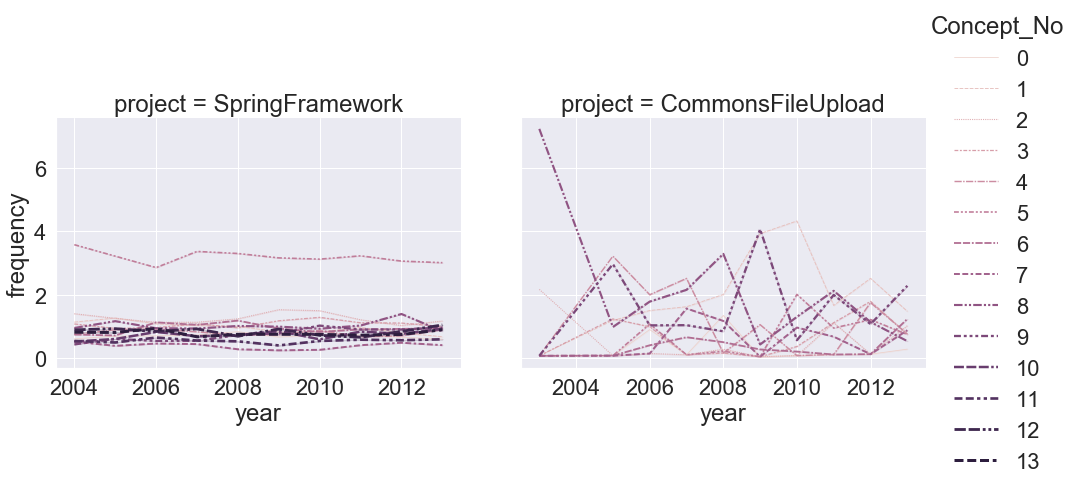}
    \caption{project duration = 10 years}
  \end{subfigure}
  
%   \begin{subfigure}{7cm}
%     % \centering
%     \includegraphics[width=5cm]{AvgWeightedMSEsImp.png}
%     \caption{Issue | Comment-level concept MSE}
% %   \label{AvgMSE}
%   \end{subfigure}

 \caption{Projects with min and max variance of the frequency of concepts at the issue-level}
 \label{MinMaxVar}
\end{figure*}

From Fig.~\ref{meanvar}, we observe that projects in each age bracket exhibit various degrees of concept evolution. %Fig.\ref{MinMaxVar} depicts the projects with the lowest and the highest variance for each age bracket.
Hence, for example, measuring a 75th-percentile variance value of 0.15 for a particular project, we cannot infer what age this project might be. Yet, we observe that younger projects tend to come with a slightly lower variance than older projects. Overall, we notice that projects with a duration of 7 years have concepts that vary the most while projects of age 3 have concepts that vary the least.
To give another insight into the concept evolution, we provide the issue-level concept frequency values for the most stable and the most volatile projects of ages 3, 6, and 10 in Fig.~\ref{MinMaxVar}.
Given the stable projects (similar behavior observable also in other age brackets but not shown due to page restrictions) we notice that the majority of concepts are of the roughly equal frequency with the occasional ``core'' concept exhibiting higher but stable frequency.

%\subsubsection{Fluctuating vs stable projects}
%In each project age bracket, we observed the mean variances at both issue-level and comment-level. 
%Fig.~\ref{MinMaxVar} illustrates for each age bracket the projects with minimum and maximum 75th-percentile variance in the frequency of issue-level concepts. The visualization shows that the frequency of concepts keeps on increasing and decreasing over time for the project with maximum variance. 
% This could be due to many factors such as employee turnover, new comers joining the team, or the arrival of issues with similar concepts.
%We also observed that for projects with minimum variance, the majority of concepts are of roughly equal frequency with the occasional ``core'' concept exhibiting higher but stable frequency (see, for example, ~Fig.~\ref{MinMaxVar}[(a), (b) and (c)]. %  the frequency of concepts does not change over time for the projects with minimum variance.

%compared the issue-level concepts' distribution against each developer over time and measured how the comment-level concepts change with regards to the issue-level concepts. In order to formalize the process of answering RQ-2, we analyzed the frequency of a single concept and compared both the issue-level and the comment-level in order to understand the correlation among both. This will help us understand how many developers share the concept which has also appeared at the issue-level.

\subsection{Developer (comment-level) and issue-level concept frequency alignment}
% To address RQ-3 ``How does concept frequency at the developer level change in comparison to the issue-level?'',
We further investigated to what extent comment-level concept frequency deviates from issue-level frequency, and whether changes at issue-level come with similar strong changes at the comment-level.

\begin{figure}[!htbp]
  \centering
  \includegraphics[width=0.8\linewidth]{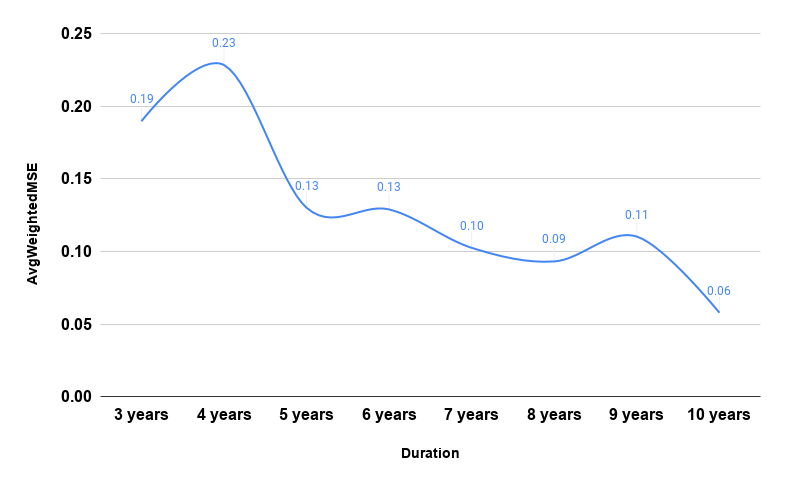}
  \caption{Issue | Comment-level concept MSE}
  \label{AvgMSE}
\end{figure}

In order to evaluate whether the comment-level concepts align with the issue-level concepts, we calculated for every developer (comment) and issue-level concept frequency, the mean squared error (MSE) of the issue-level concept frequency, and the developer-specific concept frequency. %MSE between the frequency of concepts at both levels which will indicate how far the comment-level concepts are from the issue-level concepts as the project matures. 
We then take the mean MSE across all projects of the same age bracket as an indicator of whether developer (comment) to issue-level concept alignment shows some project age-based trend. 
Fig~\ref{AvgMSE} shows these mean MSE values. From the graph, we observe that mean MSE decreases as the projects get older, hence an increasing alignment of comment-level concept frequency with issue-level concept frequency. This phenomenon could be the result of more developers becoming increasingly familiar with larger areas of the project and thus resulting in an increasingly shared domain knowledge.

An important aspect to gain insight into is whether there exists a small group of developers in OSS projects that possess the most knowledge of an important concept. We call these developers as keepers. When these keepers (or a non-negligible subset) are to leave, a significant amount of domain knowledge could be lost. 
To determine the set of keepers, we select for each project and year the most important issue-level concept and select all developers active in that year. We then sort all developers in descending order by their normalized frequency of that concept (recall the sum of frequency for a concept equals one). We then count for how many developers we need to sum up their weight to achieve an arbitrary threshold (here 0.5). The lower this threshold is set, the smaller the set of keepers will become. As the keeper count increases, we no longer would consider such a set of developers to be actual keepers but rather to describe well-distributed concept familiarity.

\begin{figure}[!htbp]
  \centering
  \includegraphics[width=0.8\linewidth]{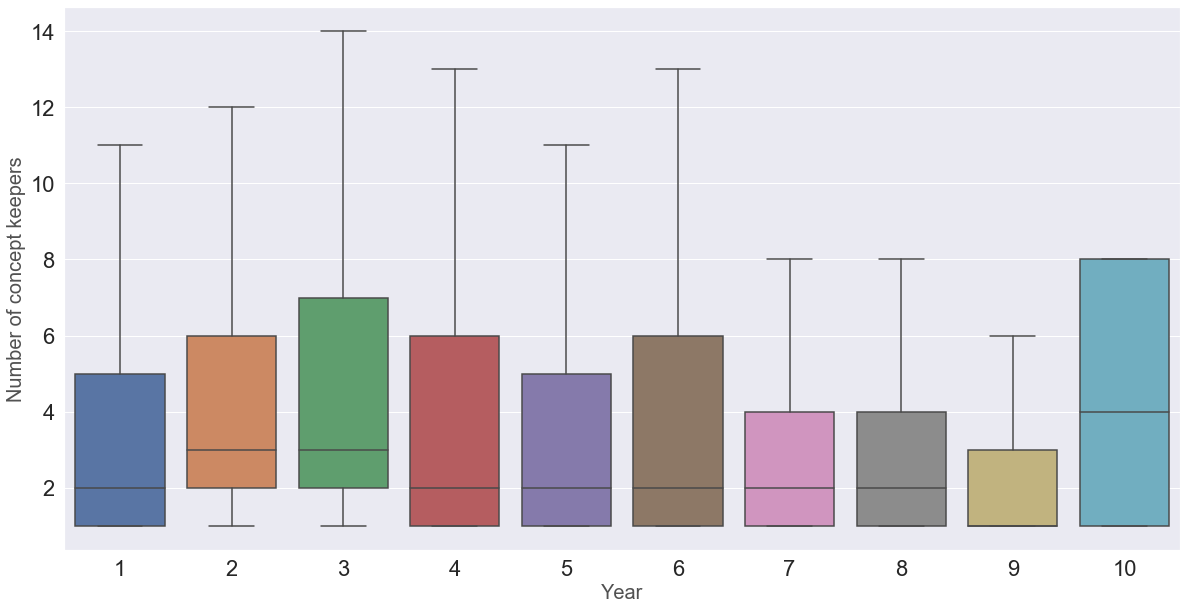}
  \caption{Amount of keepers for the most important concept per year across all projects (without outliers).}
  \label{HighAssoc}
\end{figure}

Fig.~\ref{HighAssoc} displays for each year of a project's duration the number of keepers for the most important concepts (at a threshold of 0.5) as a boxplot. Overall the boxplot shows that a single, two, or three keepers are quite common across all project years (i.e., see the median). 
Hence, even mature projects (esp. in their 7th, 8th, or 9th year) that would have had time to distribute know-how, are prone to have a single or two keepers for their most important concept (i.e., the median is 2 or lower).  
In contrast, keepers tend to be less prominent in a project's third year.

%We also analyzed the number of key developers in these projects who possess concepts that were of the highest frequency for a particular year. This will allow us to identify the ratio of key developers over the years and their evolution as the project matures. 

%Fig.~\ref{HighAssoc} shows the range of this count for the various age brackets. %the developer-level concept frequency associated with the concepts having the highest issue-level concept frequency. 
%Based on these numbers, we observe that project with shorter duration (3-6 years) tend to have a small set of keepers. Whereas, projects with longer duration (7-10 years) tend to have a larger number of developers associated with the most important concepts. In these cases, we can no longer consider these developers as ``keepers''. Note, however, that in every age bracket, there exist projects that have one or a few keepers, thus relying strongly on these developers to remain with the project.

% ////////////////// REVISED BY CMD UNTIL HERE 

% \begin{center}
%     \fbox{
% 	\begin{minipage}[c]{0.95\linewidth}
% 		\textbf{Summary of RQ-2:} The frequency of concepts change over the course of software development. However, slightly higher variance appears to exist in older projects. % (i.e., 7, 8, 9, 10 years).
% % 		We hypothesize that this phenomenon is (on average) due to short duration projects focusing primarily on their core idea and refining it, while long duration projects are on average more prone to have matured concepts (thus requiring less focus and ceasing in frequency) and have new concepts emerge that represent new needs.  
		
% 		% which implies managers should put more emphasis on balancing the domain knowledge within the team as the project matures.
% 	\end{minipage}
%     }
% \end{center}

\begin{center}
    \fbox{
	\begin{minipage}[c]{0.95\linewidth}
		\textbf{Summary of RQ-2:}  
% 		We find no correlation between the issue-level and comment-level concept frequency changes. 
		The frequency of concepts changes over the course of software development. However, a slightly higher variance appears to exist in older projects. Furthermore, a small set of keepers is prevalent in the majority of projects regardless of their project age. This observation is in line with an open-source project exhibiting a small set of core developers. %Both the analysis of developer concept frequency deviation from issue-level concept frequency as well as the keeper analysis show that long duration projects overall tend to exhibit higher spreading of concepts across developers than the shorter ones, thus indicating a higher resilience to developer turn over. Yet, we find projects with a low number of keepers in every age bracket. % Moreover, projects with longer duration have shown less MSE value indicating a strong alignment of concepts at both levels as the project evolves.
	\end{minipage}
    }
\end{center}

%We chose to address RQ-4 ``How does concepts frequency changes if a developer leaves?'' 
%at the end as the initial analysis performed to address RQ-2 and RQ-3 is mandatory to form basis in terms of change in concepts prior to discussing developer turnover. 
\subsection{Measuring the effect of leaving members}

To answer RQ-3 ``Are concept metrics able to measure the effect of a leaving developer on team knowledge?'' we first need to identify leaving members, then determine their prior absolute concept frequency, and subsequently measure the concept frequency upon their departure.

%Results of the previous RQs have shown that there is not much difference in concept frequency change and comment to issue-level correlation across the different project age groups. Hence, for this analysis, we report our findings for all projects together for sake of space and clarity.

We identify a leaving developer based on their activity level in terms of comments. To this end, we count the number of comments of a developer for each quarter (thus splitting each year into four 3-month time windows). We then tag a developer as having left in quarter $q_t$ if their number of comments in $q_t$ is lower than 10\% of the average comment count across the prior four quarters ($q_{t-1}$ to $q_{t-4}$) and stays that low for the subsequent three quarters ($q_{t}$ to $q_{t+3}$). 
Choosing quarters as the time window size strikes a balance between insensitivity to regular periods of lower activity such as vacation time and accurately pinpointing a developer's actual departure. Note that our definition allows developers to remain present within the project but merely at a very low activity level, hence having a similar effect as a developer that has actually left the team.
Applying this threshold to our data set identified 456 developers that exhibited a sharp drop in commenting activity. 

%RQ3 - expecting a drop if keeper leave (common sense) but need to know the least supported concept have least drop (results)

One would expect that a leaving developer has a negative effect on concept frequency, most so on the concept that the developer is most familiar with (their strongest concept as measured by their commenting activity). %This negative effect might be compensated by other team members which would required further investigation. In addition, we are also interested in knowing how a less supported concept is effected later on in the project?

We need to ensure that the concept frequency changes (due to the leaving developer) are caused by the concept distribution and not just by the shift in commenting behavior (i.e., the remaining team members taking over). Especially when the leaving developer is one of the core developers of the project we might see a drop in concept frequency across all concepts independent of concept distribution. 
Indeed, when we measure the rank of the leaving developers by calculating their rank based on the number of comments they made during $q_{t-1}$ to $q_{t-4}$ we find that most of the leaving developers are found within the top 10. %This means that these developers were making the most comments on issues in the past.

Hence, to measure the effect of the concept distribution, we measure if a concept's absolute frequency drops more or less than the median absolute concept frequency. 
Specifically, we calculated the absolute concept frequency, i.e., $\textit{acf(c)} = \sum_{i=1}^{n}w_{\mathit{c}}(n)$ for each concept c over all comments regardless of the developer, and \textit{acf(d,c)} for the frequency of concept c when only considering developer d over the time windows before and after a developer left, i.e., $q_{t-1}$ to $q_{t-4}$ and $q_{t}$ to $q_{t+3}$, respectively.
We thus obtain $freq_{pre,c}$ and $freq_{post,c}$ and then determine the increase or decrease in concept frequency by taking the difference, i.e., $diff_c = freq_{post,c} - freq_{pre,c}$. 

\begin{figure}[htbp]
  \centering
  \includegraphics[width=0.8\columnwidth]{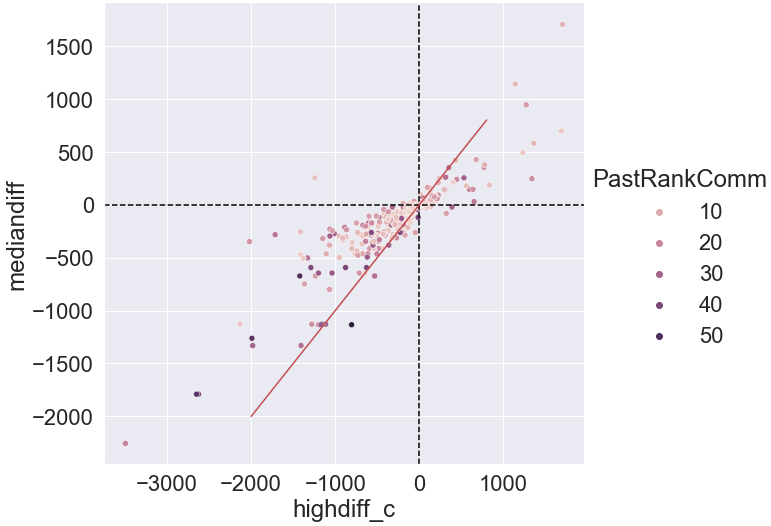}
  \caption{Comparison of leaving developer's strongest concept frequency change to median concept frequency change}
  \label{HighCorrScatter}
\end{figure}

\begin{figure}[htbp]
  \centering
  \includegraphics[width=0.8\columnwidth]{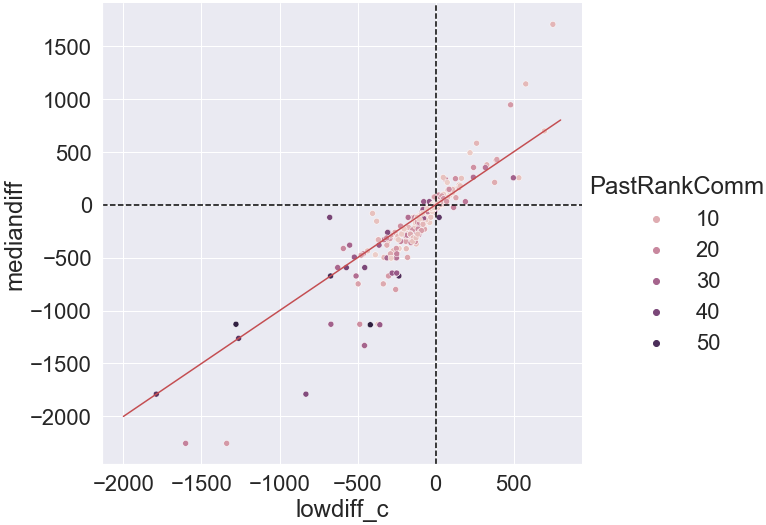}
  \caption{Comparison of leaving developer's weakest concept frequency change to median concept frequency change}
  \label{LowCorrScatter}
\end{figure}

For a particular concept, we can then plot the impact of a leaving developer in terms of change in absolute concept frequency compared to the median concept frequency change. 
We did this in Fig.~\ref{HighCorrScatter} and Fig.~\ref{LowCorrScatter} for the strongest and weakest concepts of the leaving developer.
For Fig.~\ref{HighCorrScatter}, we identified the concept the leaving developer was most familiar with based on their absolute concept frequency $max ( acf(d,c) )$ from the prior four quarters.
We then printed the difference $diff_c$ for this concept in the scatter plot. Likewise, we printed the values for the concept the developer was least familiar with (i.e., $min ( acf(d,c) )$) in Fig.~\ref{LowCorrScatter}.

The red diagonal line indicates the situations where the concept frequency change equals the median frequency change. A data point below the diagonal in the negative range describes an under-proportional drop in concept frequency while a data point above the diagonal in the negative range describes an over-proportional drop in concept frequency. In the positive range, a value under the diagonal indicates an over-proportional increase in concept frequency.
 
If concepts are equally distributed across team members, we would see the data point in both scatter plots roughly equally distributed on and around the diagonal.
Fig.~\ref{HighCorrScatter} and Fig.~\ref{LowCorrScatter}, however, clearly show a different behavior. From Fig.~\ref{HighCorrScatter} we observe for most data points that the concept frequency of the concept the developer was most familiar with dropped stronger than the median frequency change. For the least familiar concepts (in Fig.~\ref{LowCorrScatter}), we see a less severe drop in absolute concept frequency for most data points.
Interestingly, for the minority of data points where the absolute concept frequency increased in the time after departure, we observe a slightly inverse phenomenon: the least familiar concept does not increase as much as the median, and the most familiar concept increased more than the median. Further investigations are needed to understand whether the concept distribution in these few team contexts is different from the situations where a leaving developer leads to a drop in median absolute concept frequency or whether other factors can explain this result. To this end, the amount of commenting the leaving developer did as measured by the past comment rank (indicated by a data point's color in Fig.~\ref{HighCorrScatter} and Fig.~\ref{LowCorrScatter}) seems to make no impact.

\begin{center}
    \fbox{
	\begin{minipage}[c]{0.95\linewidth}
		\textbf{Summary of RQ-3:}  
		We find that a leaving developer's effect on the concept frequency is not explained by the number of comments but by the developer's concept frequency. In most cases, the developer's most familiar concept will experience an over-proportional drop in frequency, while the developer's least familiar concept will experience an under-proportional drop.
	\end{minipage}
    }
\end{center}

% EXTENSION IDEAS: 
% - ensure developer really left for whole 1 year, not just 1 quarter
% - separate between any concept and most important team concept
% - take every concept from leaving member
% - cross check with team level evolution: how many of those changes can be attributed to leaving devs?

\subsection{Knowledge distribution in OSS projects}

To address RQ-4: Is there a difference in project/concept distribution for those developers in the negative quadrant compared to the positive quadrant? We first examined the $diff_c$ of the developer's absolute concept frequency $max ( acf(d,c) )$ with respect to the median concept frequency change as illustrated in Fig.~\ref{HighCorrScatter}. As part of the observation, we then analyzed the differences in projects with data points appearing in the negative and positive quadrants. 25 projects had data points in both the quadrants leaving 132 projects with data points only in the negative quadrant, whereas 4 projects with data point only in the positive quadrant.

\begin{figure}[htbp]
  \centering
  \includegraphics[width=0.8\columnwidth]{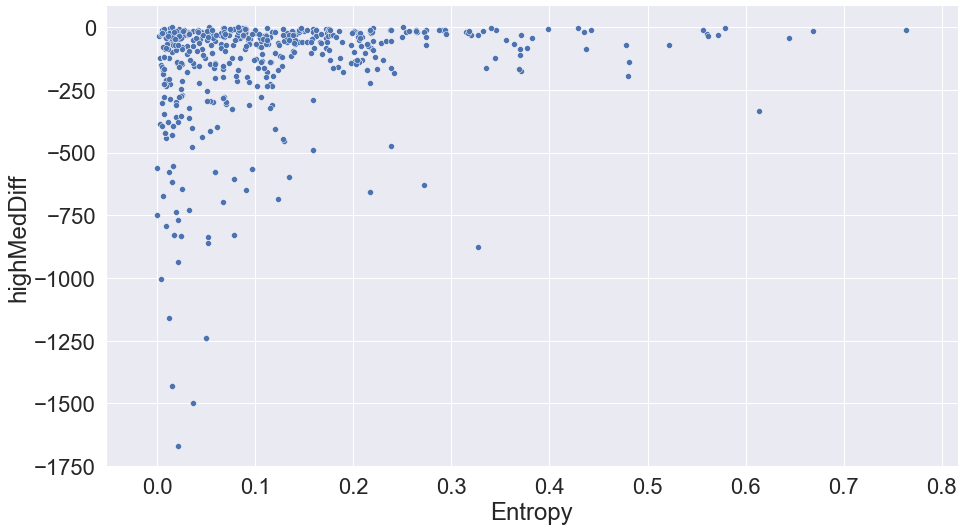}
  \caption{Entropy vs $HighMedDiff_Neg$ (OSS projects)}
  \label{Entropy-HighMedDiff_Neg}
\end{figure}

\begin{figure}[htbp]
  \centering
  \includegraphics[width=0.8\columnwidth]{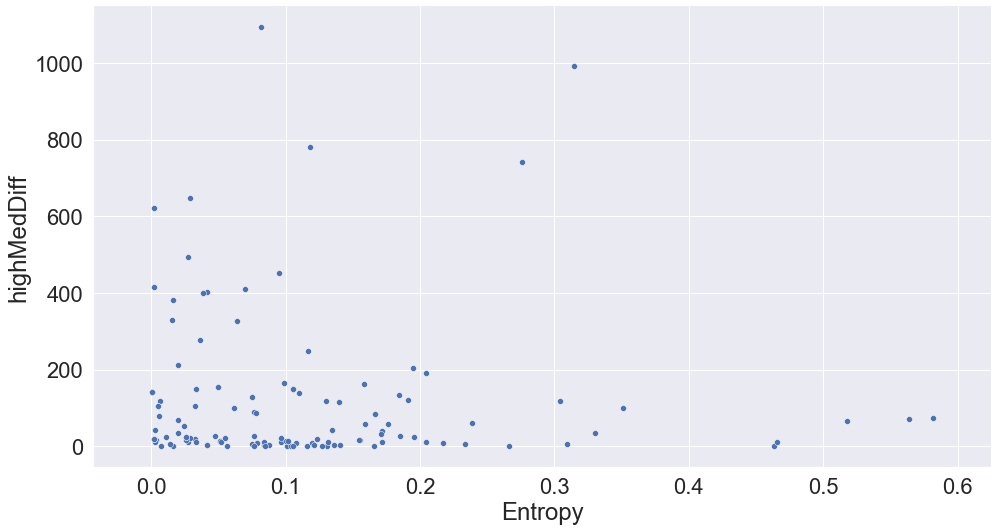}
  \caption{Entropy vs $HighMedDiff_Pos$ (OSS projects)}
  \label{Entropy-HighMedDiff_Pos}
\end{figure}

Due to the minimal number of projects appearing in the positive quadrant, we decided to investigate developer-wise concept distribution. We calculated the difference of the $diff_c$ of the developer's absolute concept frequency $max ( acf(d,c) )$ with the median concept frequency change such as $max ( acf(d,c) ) - median change$ which showed us how far on average the strongest concept of the developer is from the team in terms of frequency. We then calculated the Entropy of the top 5 developers' frequencies in the project which essentially indicates the equality of the distribution among the team. Where an entropy of 0 indicates an unequal distribution of concepts in the project while 1 represents an equal distribution. Ultimately, we draw the plots~\ref{Entropy-HighMedDiff_Neg} and~\ref{Entropy-HighMedDiff_Pos} to see the correlation between the entropy and the difference between strongest/weakest concept frequency change and median concept frequency change. Apparently, in Fig.~\ref{Entropy-HighMedDiff_Neg}, we found that the entropy is considerably lower as the diff decreases in the negative quadrant which implies that concepts are mostly not equally distributed among the teams with lower change with some exceptions. In the future, we will investigate the data points that are in the region of entropy beyond 0.5.

Looking at the Fig.~\ref{Entropy-HighMedDiff_Pos}, we see a similar behavior, however, there is a comparatively low number of data points in the positive quadrant which implies that the difference between strongest/weakest concept frequency change and median concept frequency change can only serve as one factor for the distribution of knowledge among the teams as other factors might also be the reason for influencing the Entropy.

% analyzed that the strongest concept of developers in the negative quadrant also aligns more with the highest issue-level concept as compared to the developers in the positive quadrant. This is mainly due to the higher number of developers being in the negative quadrant are the keepers of the highest issue level concepts which also corroborates the significant drop in their concept when they depart.

% We also looked into the ranks of developers based on their strongest concept, we observed that projects from the negative quadrant have lower developer ranks as compared to the projects that have data points in both quadrants. Fig.~\ref{distComparison} shows the ranks of developers of projects that have data points in both quadrants, negative quadrant, and the positive quadrant.

\begin{center}
    \fbox{
	\begin{minipage}[c]{0.95\linewidth}
		\textbf{Summary of RQ-4:} Concepts in most OSS projects do not seem to be distributed in equal proportions among the team suggesting a lack of knowledge familiarity. One reason for that could be the team factor as the different teams might not usually share knowledge with each other.

	\end{minipage}
    }
\end{center}

\subsection{Implications for recommendation algorithms}

% Recommendation ( top3/higher than top3 )

Addressing RQ5: What are the implications for the assignee recommendation algorithm that should also result in more evenly distributed knowledge? A naive intuition would be to recommend developers based on their past concept familiarity with the issues. To see whether this approach holds to the existing dataset, we divided projects into unequal and equal distribution groups, i.e., for unequal projects if max\_gate > median + 0.01, and for equal projects, if max\_game < median - 0.01. We hypothesize that developers that are less familiar with concepts should be recommended for projects in which concepts are unequally distributed while top-ranked developers in terms of concept familiarity should be recommended for equally distributed projects. Then, we rank each developer based on concept frequency meaning developers with higher concept frequency will have lower ranks.

% To evaluate the differences in ranks in both groups, for each concept row in the test set, we selected top 5 developers from the train set that match this concept. Then, we observed whether this whether assignee in the test set appear in the top 5 selected developers. This will provide us with the information when top ranked developers usually assigned to projects in which concepts are equally/unequally distributed.

We calculate mean reciprocal ranks (MRR) with values ranging from 0 to 1. Where 1 indicates devs with higher concept frequency are assigned to issues while 0 indicates devs with lower concept frequency are assigned to issues. Table~\ref{naiveapproach} shows the mean reciprocal ranks for each split in both groups. For equally distributed projects, we observed that the mean reciprocal rank is lower than unequally distributed projects which implies that developers with low familiarity with the concepts are typically assigned to new issues in equally distributed projects as opposed to unequally distributed projects. This strengthens our assumption that projects with equally distributed concepts tend to assign developers who are generally less familiar with the concepts while projects with unequally distributed concepts tend to assign higher-ranked developers thus resulting in an imbalanced distribution of concepts which leads to the dependency on the keepers.

% -- TAKE TOP 5 (RANDOMIZE LIST)

% Then, we evaluated the recommendations using the ground truth test set.

\begin{table}
\centering
\begin{tabular}{cccc}
\toprule
Type & Split & Mean reciprocal rank \\ 
\midrule
Equal & all & 0.14896954002884027 \\ 
Unequal & all & 0.368514437643808\\
%  \rowcolor{green}
\bottomrule
\end{tabular}
\caption{Naive approach}
\label{naiveapproach}
\end{table}

% ML APPROACH

% % COMPARE GROUPS
% Next, we try to explore whether the predictive models can help recommend the suitable developers for each group in this context.
% Considering this task as a supervised learning problem, we have used the tuple ($distance to the max concept dev$, freq, W) as features for the training of a machine learning (ML) model, where freq refers to concept frequency, and W represents the weight of the concept. We further used the 'Assignee' attribute as the target label. We then evaluated the prediction capability of \textit{ConceptRealm} using a benchmark study consisting of 7 alternate ML models. The choice of ML models is based on the common ML models used for multi-class classification problems in literature.

% Fig.~\ref{modelResults} shows the performance of the ML models employed in the benchmark study.

% \begin{figure}[htbp]
%   \centering
%   \includegraphics[width=0.8\columnwidth]{modelResults.png}
%   \caption{Model results}
%   \label{modelResults}
% \end{figure}

\begin{center}
    \fbox{
	\begin{minipage}[c]{0.95\linewidth}
		\textbf{Summary of RQ-5:}
		
		Supporting our hypothesis, we observed that OSS projects that have equally distributed concepts tend to assign developers who are less familiar with these concepts thus gaining equal distribution of concepts while in contrast, projects with unequally distributed concepts tend to assign top-ranked developers.

	\end{minipage}
    }
\end{center}

\subsection{Industrial evaluation}
Addressing RQ6: How effective \textit{ConceptRealm} can be in the detection of concepts and identifying the impact of leaving developers in the industry? We have performed a preliminary field study of our approach with the help of an industrial case study from Dynatrace and an open-ended questionnaire with a core practitioner.

\subsubsection{Data extraction}

We first prepared a python script to extract the issues' attributes from this closed-source JIRA project. We obtained project data spanning the course of six years. Before extraction, we were also required to anonymize some of the attributes to maintain the integrity of sensitive information, e.g., assignee/reporter names, user ids, team names, team roles, etc. In total, we obtained 49457 issues and 168608 comments. Issues extracted were then passed through the preprocessing and concept extraction as highlighted in the aforementioned Section~\ref{dp} and Section~\ref{CE}, respectively. While constructing~\textit{ConceptRealm}, we treated each team as an individual OSS project in order to have team-level insights. Fortunately, the difference from the OSS projects is that, with access to this closed-source project data, we also had the opportunity to observe concepts of individual teams within this project. This would allow us to generate a very granular set of concepts that can be vital in determining the high-level domain knowledge present within teams as well as individual developers. Thus, we investigated the team-based concept distribution within this project in order to be able to observe any team-related patterns.

\subsubsection{Open-ended questionnaire with the practitioner}

Additionally, we prepared an open-ended questionnaire in line with our aforementioned research questions and communicated with the lead product manager from Dynatrace. This lead product manager has in-depth experience in the project and is directly responsible for managing the teams of this project.

We have provided a list of 10 inactive developers that were identified from our approach along with their concepts and years when they left. To reduce bias, we provided two issues for each developer to ensure accurate evaluation. This would allow us to understand whether the leaving developers are correctly identified and had previously worked on issues with similar concepts.

Questions of the questionnaire include:

\begin{center}
    \fbox{
	\begin{minipage}[c]{0.95\linewidth}
        \begin{enumerate}
            \item Q1 - Are these engineers humans and not development bots? [Yes, No, Maybe]
            \item Q2 - Are these the latest assigned engineers who are also working on similar issues fixing/implementation? [Yes, No, Maybe]
            \item Q3 - Could these engineers (their familiarity with the product) be accurately described by the concept? [Yes, No, Too generic]
            \item Q4 - Are the issues described accurately by their concept? [Yes, No, Maybe]
            \item Q5 - Did these engineers leave their team roughly in the identified year and quarter? [Yes, No, Maybe]
        \end{enumerate}
	\end{minipage}
    }
\end{center}

\subsubsection{Questionnaire results}

We are interested in knowing the practical efficacy of our approach. For this purpose, we calculate accuracy for each question as to the total number of 'Yes' provided by the practitioner divided by the total number of developers.

Findings from the questionnaire show that most of the developers (> 60\%) identified as leaving members by \textit{ConceptRealm} are valid. In addition, the practitioner confirmed that the concepts associated with the developers by \textit{ConceptRealm} are indeed aligned to the issue they are working on. Table~\ref{quesResults} shows the results of the questionnaire. The practitioner also pointed out that some concepts appear to be too generic. This is due to the fact that concepts were generated based on the entire project corpus and were not team-specific when showed to the practitioner. Nevertheless, these findings support the hypotheses postulated in this study regarding \textit{ConceptRealm} and further highlight the importance of using such an approach in practice.

\begin{table}
\centering

\begin{tabular}{c|c|c|c|c|c|c|}
%   \cline{2-7}
\toprule
\multirow{2}{*}{}                            & \multirow{2}{*}{Developers} & \multicolumn{5}{c|}{Questions}   \\ %\cline{3-7} 
&                   &  Q1 & Q2 &  Q3 &  Q4 & Q5   \\ \midrule
\multicolumn{1}{|l|}{\multirow{10}{*}{\rotatebox{90}{Evaluation}}} &        Dev 1           & \checkmark  & \checkmark &     &  & \checkmark  \\ %\cline{2-7} 
\multicolumn{1}{|l|}{}                       &       Dev 2       & \checkmark  & \checkmark & \checkmark & \checkmark  & \checkmark  \\ %\cline{2-7} 
\multicolumn{1}{|l|}{}                       &       Dev 3             & \checkmark & \checkmark & \checkmark   & \checkmark & \checkmark    \\ %\cline{2-7} 
\multicolumn{1}{|l|}{}                       &       Dev 4            & \checkmark & \checkmark &  \checkmark  &   & \checkmark   \\ %\cline{2-7} 
\multicolumn{1}{|l|}{}                       &       Dev 5            & \checkmark & \checkmark & \checkmark  &   &  \checkmark  \\ %\cline{2-7} 
\multicolumn{1}{|l|}{}                       &       Dev 6            & \checkmark & \checkmark & \checkmark  &   &  \checkmark    \\ %\cline{2-7} 
\multicolumn{1}{|l|}{}                       &       Dev 7            & \checkmark & \checkmark & \checkmark  &   &      \\ %\cline{2-7} 
\multicolumn{1}{|l|}{}                       &       Dev 8            & \checkmark &  &   &   &    \\ % \cline{2-7} 
\multicolumn{1}{|l|}{}                       &       Dev 9            & \checkmark &  &   &   &    \\% \cline{2-7} 
\multicolumn{1}{|l|}{}                       &       Dev 10           & \checkmark &  &   &   &    \\ 
\bottomrule
\end{tabular}

\caption{Questionnaire results}
\label{quesResults}

\end{table}

\subsubsection{Comparison with OSS projects}

% add same stats as compared to OSS projects

% RQ2:

\begin{figure}[htbp]
  \centering
  \includegraphics[width=0.8\columnwidth]{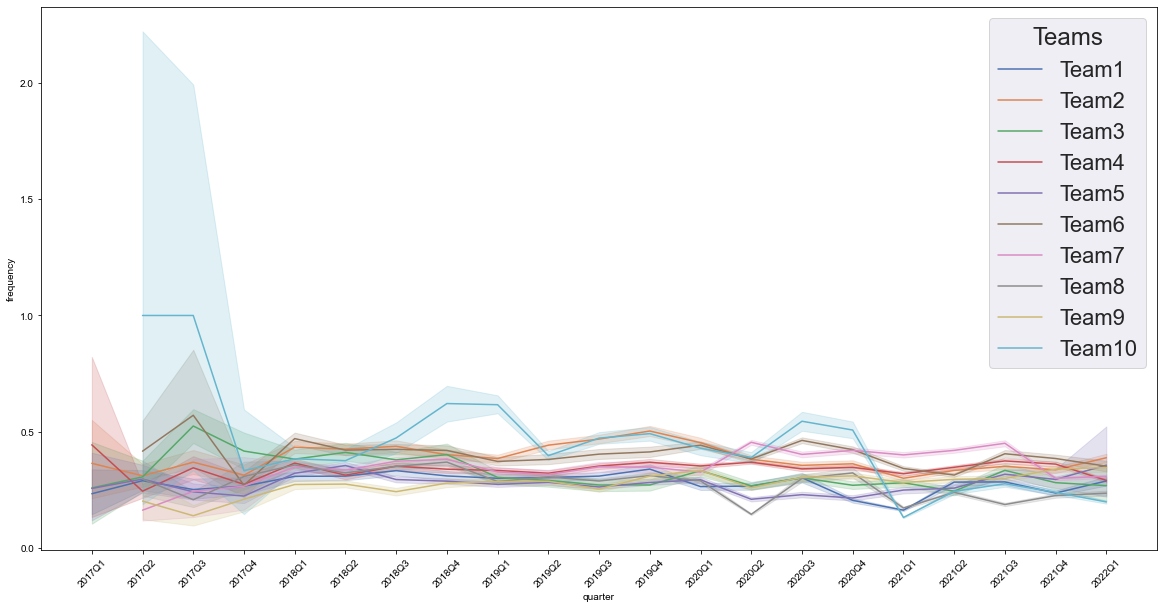}
  \caption{Team Concept Frequency Variation}
  \label{TeamConceptFrequencyVariation}
\end{figure}

Compared to OSS projects, we observe similar patterns in the evolution of the frequency of concepts for the closed-source project over the course of 6 years. Fig.~\ref{TeamConceptFrequencyVariation} shows the variation in concept frequency for the top 10 teams in the closed-source project. The concept frequency of most of the teams is apparently stable across the duration except for Team10. Further investigation on this revealed that for the year 2017, this team had only 3 active developers, thus, causing an increase in the frequency for this year.

\begin{figure}[htbp]
  \centering
  \includegraphics[width=0.8\columnwidth]{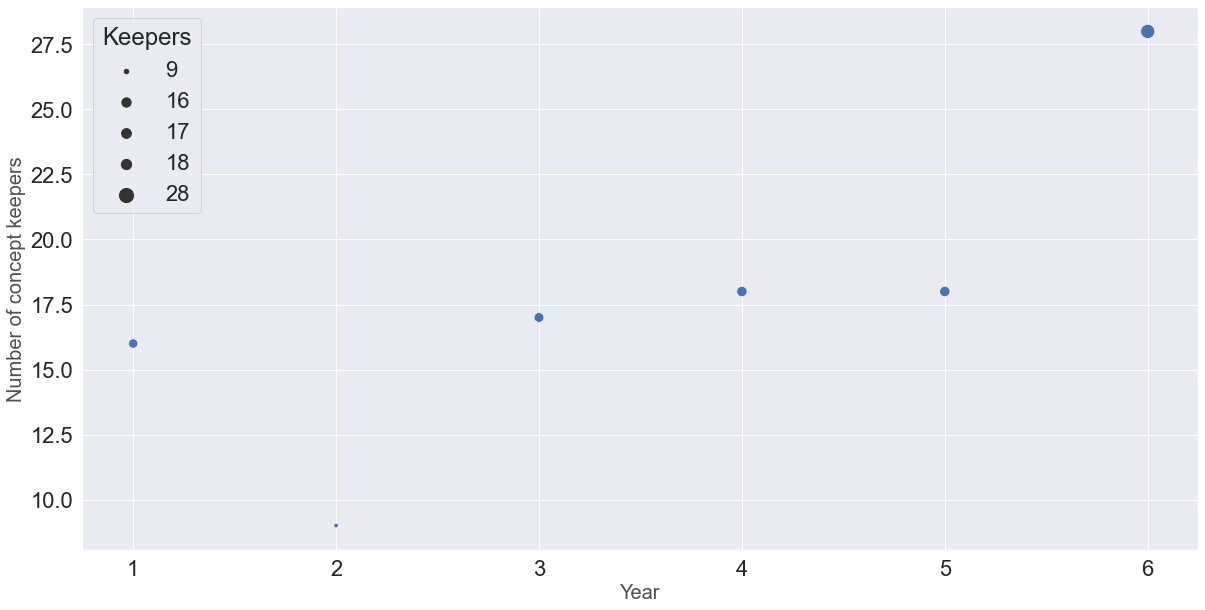}
  \caption{Dynatrace keepers}
  \label{dynatrace-keepers}
\end{figure}

Contrary to OSS projects, the closed-source project seems to have a higher number of keepers for each year. Fig.~\ref{dynatrace-keepers} shows the number of keepers for each year of the project.

% RQ3:

\begin{figure}[htbp]
  \centering
  \includegraphics[width=0.8\columnwidth]{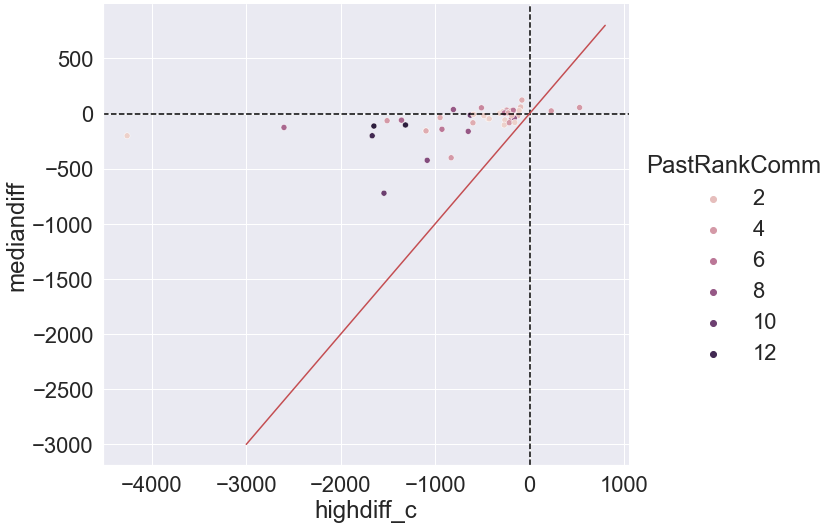}
  \caption{$MedianDiff-highdiff$ (Dynatrace project)}
  \label{MedianDiff-highdiff-dynatrace}
\end{figure}

Observing the impact of leaving developers with the strongest concepts in the team for the closed-source project, we see a similar trend as shown in Fig.~\ref{MedianDiff-highdiff-dynatrace}, i.e., the change in the frequency of developers (leaving ones) strongest concept tends to be lower than the median change in concept frequency of the team implying a higher drop in concept frequency of the leaving member.

% RQ4:

\begin{figure}[htbp]
  \centering
  \includegraphics[width=0.8\columnwidth]{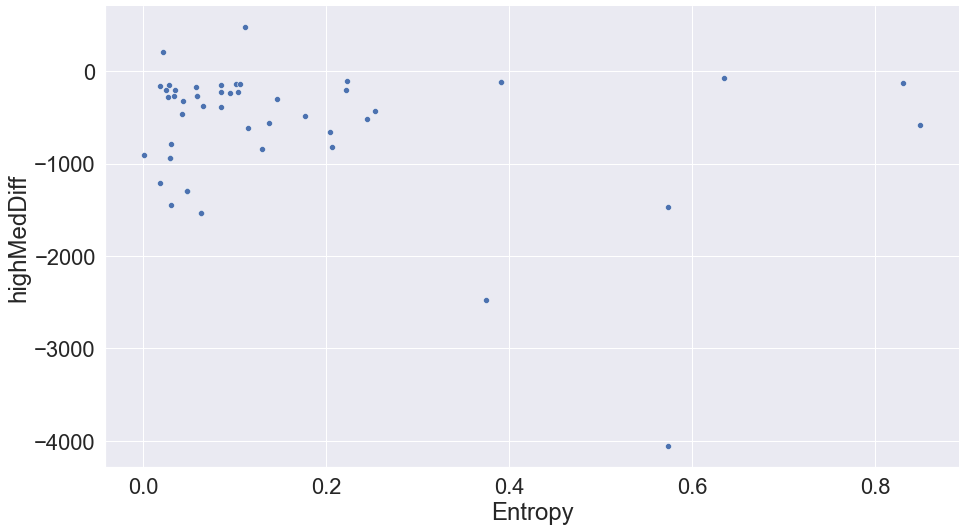}
  \caption{$highmediandiff-entropy$ (Dynatrace project)}
  \label{highmediandiff-entropy-dynatrace}
\end{figure}

To be able to better understand, we focus only on the negative quadrant. As shown in Fig.~\ref{highmediandiff-entropy-dynatrace}, we see similar behavior in the negative quadrant, the entropy is lower as the difference between strongest/weakest concept frequency change and median concept frequency decreases, which implies concepts are not equally distributed among the teams in the closed-source project as well.

\begin{center}
    \fbox{
	\begin{minipage}[c]{0.95\linewidth}
		\textbf{Summary of RQ-6:}
		
		Based on the results from the closed-source project, we observed that \textit{ConceptRealm} is able to correctly identify the keepers in the project and developers with mutual concepts are indeed aligned with the issues having similar concepts. Moreover, we see similar trends in the closed-source project as compared to OSS projects which strengthened the generalization capability and overall consistent behaviour of \textit{ConceptRealm}.

	\end{minipage}
    }
\end{center}

%% file: discussion.tex
\subsection{Discussion}
Having obtained the results for the six RQs, here we now discuss these results and their implications. % for researchers and practitioners. %The analysis of obtained results complemented with an elaboration on further implications of the \textit{ConceptRealm} is discussed here.

%The overarching goal of this study was to investigate ways to represent domain knowledge of teams in OSS projects without the need to access the source code. One challenge was to develop a representation around a developer that captures his/her knowledge of the domain as close as possible with the help of NLP. For this purpose, we focus on the communication aspect of developers to extract the developer-level knowledge and issues for issue-level knowledge and proposed the \textit{ConceptRealm} which is a form of knowledge representation of software development teams and the issues in the OSS projects.

The results of RQ1 show that our approach produces meaningful concepts as these can be applied to characterize the assignees of issues. It is not a concern that the results are not significant for all projects as other factors aside from concept familiarity typically determine an issue's assignee. Specifically, our baseline assumption that a developer working on issues of a particular concept in the past will also work on issues of that concept in the future does not necessarily have to always hold but is generally valid. The primary focus of answering RQ1 was on establishing our approach's ability to produce useful concepts, not to suggest an issue assignment metric.
Hence, we do not suggest that the developer with the highest concept frequency should be an issue's assignee but that rather concept familiarity is one important contributing factor that could be exploited for that purpose.
Rather we believe that the observation of such a correlation motivates measures to distribute concept familiarity within a developer group to enable them to work on a broader range of issues - mitigating unexpected turnover.

%, these concepts can help form a domain ontology surrounding the developers which might provide an overview of the unfamiliar concepts and perhaps motivate managers to conduct in-house training sessions in order to spread the knowledge. We also noticed that the higher the familiarity of a developer with a concept is, the more likely it is that the same developer will work on issues that exhibit similar concept.

Investigations of RQ2 have revealed that some projects have widely fluctuating concept frequency, while others are very stable, regardless of year. Across all projects, the data showed that the projects with longer duration exhibit higher variation in their concepts frequency than younger projects. We hypothesize that this phenomenon (on average) is due to short-duration projects focusing primarily on their core idea and refining it, while long-duration projects are on average more prone to have matured concepts (thus requiring less focus and ceasing in frequency) and have new concepts emerge that represent new needs. Yet, this indication of a concept evolution life-cycle is very light and we stress that the nature of the project is probably much more influential on concept evolution than its age. More investigations are needed to determine which factors exactly drive concept evolution.

Moreover, a significant amount of projects have a small number of concept keepers. Especially matured projects (i.e., project years 7 to 9) often have a single or two developers accounting for half of the familiarity with the project's most important concept. One could have expected that older projects are more likely to have distributed concept familiarity. Instead, the available data let us assume that experts, or key owners, become established that know one or multiple particular concepts very well and take on related issues. 
Finally, our analysis of abruptly leaving team members in RQ3 highlights that when these developers represent a larger share of concept familiarity of an otherwise unequally distributed concept, and that this concept is less well supported upon the developer's departure.

This is essentially the case when the developer that possesses the strongest concept becomes less engaged with the project, this strongest concept will most often see less activity in the time following their ``departure'' (as measured via comment frequency).
%seems to fade away the next consecutive year.
%Remove (we are observing a correlation in drop of concept familiarity)
This, in general, highlights the challenge of knowledge transfer in OSS projects.
% We find, however, that most such abruptly leaving developers have little gatekeeping extent and hence little effect on issue-level concept frequency. 
Combining our observations from RQ2 (many projects with keepers) and RQ3,
% (most leaving developers have low gatekeeping effect)
we conclude that keepers rarely leave the project or, if they leave, then they would not do so in an abrupt manner. This observation may be subject to survivor bias, i.e., having only ongoing/successful projects in the data set. Additional investigation on abandoned projects should yield more insights into whether one of the abandonment causes is the departure of a keeper.

To gain a more in-depth understanding  of why is this case, we observed from RQ4, that the concepts in OSS projects do not seem to be equally distributed among the developers and when they leave, a significant drop in their concept frequency is likely to follow. Reasons behind this behavior could be the lack of knowledge sharing among the developers/teams and perhaps the geographically remote nature of teams in OSS projects.

We also investigated the implications that can be inferred from~\textit{ConceptRealm} for modern-day assignee recommendation algorithms for RQ5. We separated the OSS projects into two equally and unequally distributed concept groups based on a threshold that we intuitively defined. This threshold helps us to include projects that are skewed towards the extremes of the two proportions. i.e., < median - 0.01 and > median + 0.01. We also discovered that developers who are top-ranked based on their concept frequency tend to be assigned to new issues in the projects thus making an unequal distribution of concepts. While developers that are less familiar with the concepts are mostly assigned to new issues which in turn results in equal distribution of concepts. This also strengthens our hypothesis that projects tend to have an equal distribution of knowledge when developers that have low concept familiarity are assigned to new issues.

% about the industrial project
For RQ6, we also investigated the usefulness of~\textit{ConceptRealm} by extracting the data from the closed-source project from Dynatrace. Due to easy access to the teams, we decided to perform the team-wise analysis of the project. We evaluated each research question based on each team within the Dynatrace project. We observed that the variation of concepts within teams of the closed-source project is similar to the OSS projects. Moreover, we investigated how similar the impact of leaving members is within a closed-source project as compared to OSS projects. We found a similar drop in concept frequency for strongest concept keepers when they depart the team or the project. This is essentially relevant to the assumptions we postulated for the OSS projects which apparently are also correct for the closed-source project.

Then, we conducted an open-ended questionnaire with the lead practitioner of Dynatrace. Findings from this questionnaire indicate that this approach is indeed helpful in identifying the individual that possesses the strongest concepts in the project and vice versa. Similarly, this approach also helps in understanding whether the issues are being assigned to the right person that might or might not be familiar with the concepts associated with that issue.

Compared to the OSS projects, the closed-source project is quite similar except for the number of keepers which might differ due to the size of the project. 
In essence, this approach guides the practitioner in making such development-oriented decisions and further highlights the overall team clusters present within the project.

% \vspace{-4.25mm}

\subsection{Implications}

%Implications how can the approach be applicable to industrial projects but results could differ

The primary focus of this paper is on introducing the~\textit{ConceptRealm} and demonstrating its usefulness with the example of keeper analysis. %Future work will focus more on gatekeeping properties.
From the results, we conclude that this research has important implications for the scientific community as well as practitioners.

\subsubsection{Implications for researchers} The definition of the concept realm, especially the metrics for concept frequency, allows to measure concept familiarity distribution in a team (and subsequently identify key developers in a project). Concept frequency thus constitutes another factor that may help to characterize successful projects. The~\textit{ConceptRealm} thus becomes another tool to study the assignments of developers to issues and subsequently the coordination among team members. Applying our approach to other artifacts aside from issues and their comments offers the opportunity to compare concepts extracted from requirements, documentation, or source code, and how these concepts differ from those extracted from issues. The~\textit{ConceptRealm} thus serves as another view onto a development team, and especially for open source systems, the reliance on key developers. This subsequently serves as a basis to compare against concept distribution and evolution in industry projects.

\subsubsection{Implications for practitioners}
While the \textit{ConceptRealm} is not readily integrated into a software engineering support tool, it would ultimately become an important basis for measuring concept distribution in development teams (open source and industrial). 

In contrast to OSS development, we would expect that concepts are much more distributed, i.e., shared, in industrial settings due to two main aspects: first, team members are more stable and continuously available, and second, management actively aims to reduce the impact of turnover by encouraging concept distribution. Further studies are needed to provide more insights into this aspect.

Knowing the OSS systems have volunteering members that might not be as stringent to the project as a company's contract-based employees, we believe that the proposed representation could greatly help in measuring the concept distribution of teams within these industrial projects as well but  the distribution might vary compared to OSS systems. However, further studies are required to investigate this aspect.

% this is something that needs to be investigated (the dist might vary)

The combination of the issue-level frequency with the keeper analysis allows us to identify not just any concept that is poorly spread in the team but, more importantly, identify those that are currently important for the project (i.e., at the team-level). Having only one or a few keepers of low-frequency concepts might be acceptable as a leaving keeper will have less impact than a keeper for a high-frequency concept. Along these lines, the~\textit{ConceptRealm} helps stakeholders to identify the developers with the highest familiarity with a concept and steer the assignment of new issues more towards developers that are perhaps not optimally but sufficiently suitable in order to improve the concept spreading. For newcomers, the~\textit{ConceptRealm} may help to identify the pre-existing concepts and who is familiar with them to more effectively identify the right person for questions, reviewing, or bug reports. The~\textit{ConceptRealm} can be leveraged to identify and estimate the impact of leaving keepers and prioritize concepts that need to be better distributed.

%% file: ThreatsToValidity.tex
\textit{External validity}:
% This threat refers to the employment of a below-par research methodology.
%Moreover, we have followed a standardized and systematic way to perform each activity supported by the literature. We also made sure that there is no bias in 
We address researcher bias by relying on a large data set of real open source projects of non-negligible size. 
While the data source was limited to the issue tracker Jira, this study was not specific to Jira as any textual source from an issue tracker can be used, and the use of issue trackers is very common nowadays with little difference among the popular trackers respective to the extracted data (i.e., issue description and comments).
While other sources such as discussion lists may also provide useful information where issue comments are less intensively used, previous research has shown that these serve similar purposes~\cite{Panichella2014} and hence could be used as a substitute data source.

\textit{Internal validity}:
Also, we aimed to avoid introducing bias in identifying the number of concepts for each project. To this end, we performed a sanity check as outlined in Section~\ref{SC} with the construction of a sufficiently large number of LDA models, measured the overlap and cohesion of the obtained concepts, and selected the number of optimal concepts yielding the highest cohesion and lowest concepts overlap.

\textit{Construct validity}:
One key study design decision was to consider only issues and their comments (rather than also/instead of considering source code, requirements, or documentation, etc.). On the one hand, this allows to include also team members that are not necessarily contributing source code, and, at the same time, allows to observe concept changes over time without having to tediously extract the exact changes a developer made to a particular artifact at a particular time (hence remaining also programming language agnostic). While we haven't analyzed to what extent concepts might emerge differently when including source code, we restricted our evaluation to issue-centric aspects such as assignment rather than, e.g., pull request reviewing.

% \subsection{Construct validity}
% As we are providing a representation of developer-issue interaction through concepts extracted from issues and their corresponding comments. The goal of this work is to provide a representation for the problem space where the evolution of domain knowledge can be measured by analyzing the transitions in concepts appearing within a team and how they are changing over time with respect to various factors. Therefore, the data extracted from issues and comments is sufficient for this study.

% \subsubsection{External validity}
%It is noteworthy to mention that the overarching goal of this study is to propose a novel representation which is capable of capturing the high-level domain knowledge distribution in a software development team. Like many other studies, this study is also prone to certain threats. One which refers to the generalizability of the approach with respect to the availability of scarce data. In order to overcome this threat, we have employed a sufficiently large dataset comprised of more than 500 projects having over 300k issues, and over a million comments, thus strengthening the conclusions made in the study.

\textit{Conclusion validity}:
The findings of this study, as the title highlight, apply primarily in the context of open-source software development. Hence, we cannot conclude that industrial projects experience similar levels of concept evolution and similar levels of keepers. Our approach, however, should be well applicable to industrial software development contexts where a significant amount of know-how is captured in issues and their comments. Contexts, where most interaction among developers is occurring face to face, might not benefit from our approach as the extracted developer-centric concept frequency values are likely to not accurately reflect the team's actual concept familiarity distribution.

%The results of this study should be carefully interpreted. The definition for the concepts representing the domain knowledge is based on logical assumptions and practicality. We only studied OSS projects therefore our conclusions are only valid for OSS software projects. Further studies are needed to enhance the capabilities of capturing the domain knowledge for industrial projects.

%Although the \textit{ConceptRealm} representation is based on the textual features present in the issues, similar vectorization can also be employed to other attributes contributing to domain knowledge such as developer commits and pull requests.

%% file: conclusion.tex
This paper analyzed how concepts evolve within OSS development teams. For this purpose, we constructed a practical and general representation of domain knowledge denoted as the \textit{ConceptRealm}, which characterizes a developer's concept familiarity extracted from the developer's involvement in issue tracking systems. Our analysis of OSS projects shows that our approach produces valuable concepts that can be applied, for example, to characterize future developer-issue associations. Another application of the \textit{ConceptRealm} is for investigating (and ultimately estimating) the impact of leaving team members. For example, we found that a concept's keeper that leaves the team will likely result in a subsequent drop in the frequency of the respective concept. Furthermore, we also evaluated~\textit{ConceptRealm} using an industrial case study which attested to the usefulness of this approach. We believe this representation allows managers better to align issues with the developers' concept familiarity and helps identify key individuals in the team. The \textit{ConceptRealm} could thus serve as the basis of novel recommendation systems.

Having established the basis for measuring concept familiarity in a team, we intend to focus more on applying these metrics to investigate in more detail the impact and role of keepers in OSS projects and compare these findings to teams and their concept distribution evolution in larger closed-source systems. This in-depth keeper analysis will also focus on their role in abandoned projects.